\documentclass[10pt]{article}
\pdfoutput=1
\usepackage{jheppub,amsmath,amssymb}
\usepackage{enumerate}
\usepackage{bm}
\usepackage{bbm}
\usepackage{enumitem}
\usepackage[usenames,dvipsnames]{xcolor}
\usepackage{amsmath}
\usepackage{amsfonts}
\usepackage{amssymb}
\usepackage{graphicx}
\usepackage{hhline}
\usepackage{subfig}
\usepackage{hyperref}
\usepackage{color}
\usepackage{verbatim}
\usepackage{amsmath,amsthm,amssymb}
\usepackage{lscape}
\usepackage{float}
\usepackage{caption}
\usepackage{lscape}
\usepackage{breqn}
\usepackage{multicol}
\usepackage{graphics}
\usepackage{tikz,todonotes}
\usepackage[utf8]{inputenc}
\usepackage{autobreak}
\usetikzlibrary{arrows,positioning,decorations.markings,decorations.pathmorphing,calc}
\pgfdeclarelayer{edgelayer}
\pgfdeclarelayer{nodelayer}
\usetikzlibrary{decorations.pathreplacing}
\pgfsetlayers{edgelayer,nodelayer,main}
\tikzset{none/.style={draw=none}}
\tikzset{new edge style 2/.style={black}}
\tikzset{new style 0/.style={black}}
\tikzset{rednode/.style={draw=none, scale=0.3pt,fill=red,circle, draw}}
\tikzset{redline/.style={line width=0.3mm,red}}
\tikzset{greyE/.style={line width=0.1mm,gray}}
\usepackage[utf8]{inputenc}
\usetikzlibrary{arrows,positioning,decorations.markings,decorations.pathmorphing,calc}

\definecolor{hyperref}{RGB}{026,028,087}

\newcommand{\beq}{\begin{equation}}
\newcommand{\eeq}{\end{equation}}
\newcommand{\bea}{\begin{eqnarray}}
\newcommand{\eea}{\end{eqnarray}}
\def\be{\begin{equation}}
\def\ee{\end{equation}}

\def\beq{\begin{equation}}
\def\eeq{\end{equation}}

\newcommand{\M}{M}
\newcommand{\K}{\mathcal K}
\newcommand{\U}{\mathcal U}

\renewcommand{\[}{\left[}

\renewcommand{\L}{\mathcal L}

\def\be{\begin{equation}}
\def\ee{\end{equation}}
\def\ba{\begin{eqnarray}}
\def\ea{\end{eqnarray}}

\def\nn{\nonumber}

\def\d{\mathrm{d}}

\usepackage[normalem]{ulem}

\def\ba{\begin{eqnarray}}
\def\ea{\end{eqnarray}}

\def\L{\mathcal{L}}
\def\K{\mathcal{K}}

\def\E{\mathcal{E}}

\def\stu{St\"uckelberg }
\newcommand{\Lic}{{Lichnerowicz }}

\def\d{\mathrm{d}}
\def\mn{_{\mu \nu}}
\def\ab{_{\alpha \beta}}
\def\mupn{^\mu_{\, \nu}}
\def\({\left(}
\def\){\right)}

\def\p{\partial}
\def\ie{{\em i.e. }}

\def\C{\cos \theta}
\def\S{\sin \theta}

\begin{document}

\title{Positivity Constraints on Interacting Spin-2 Fields}

\author[a]{Lasma Alberte,}
\author[a,b]{Claudia de Rham,}
\author[a]{Arshia Momeni,}
\author[a]{Justinas Rumbutis,}
\author[a,b]{Andrew J. Tolley}
\affiliation[a]{Theoretical Physics, Blackett Laboratory, Imperial College, London, SW7 2AZ, U.K.}
\affiliation[b]{CERCA, Department of Physics, Case Western Reserve University, 10900 Euclid Ave, Cleveland, OH 44106, USA}

\emailAdd{l.alberte@imperial.ac.uk}
\emailAdd{c.de-rham@imperial.ac.uk}
\emailAdd{arshia.momeni17@imperial.ac.uk}
\emailAdd{j.rumbutis18@imperial.ac.uk}
\emailAdd{a.tolley@imperial.ac.uk}

\abstract{The consistency of the EFT of two interacting spin-2 fields is checked by applying forward limit positivity bounds on the scattering amplitudes to exclude the region of parameter space devoid of a standard UV completion. We focus on two classes of theories that have the highest possible EFT cutoff, namely those theories modelled on ghost-free interacting theories of a single massive spin-2 field. We find that the very existence of interactions between the spin-2 fields implies more stringent bounds on all the parameters of the EFT, even on the spin-2 self-interactions. This arises for two reasons. First, with every new field included in the low-energy EFT, comes the `knowledge' of an extra pole to be subtracted,  hence strengthening the positivity bounds.
Second, while adding new fields increases the number of free parameters from the new interactions, this is rapidly overcome by the increased number of positivity bounds for different possible scattering processes.  We also discuss how positivity bounds appear to favour relations between operators that effectively raise the cutoff of the EFT. }

\maketitle


\section{Introduction}

In a recent work \cite{Alberte:2019lnd} we considered two classes of effective field theory (EFT) descriptions for interactions of multiple spin-2 fields \cite{Hinterbichler:2012cn}, each based on the requirement that the cutoff of the effective field theory was as high as possible. These EFTs were constructed using the assumption that the symmetry that is spontaneously broken by the mass term is diffeomorphism invariance. This assumed symmetry breaking imposes a particular organizational structure for the EFT expansion based on power counting derivatives and interactions of the associated \stu fields (Goldstone modes). The assumed diffeomorphism symmetry breaking implies that the interactions of the spin-2 states are naturally built nonlinearly out of a vierbein or metric variable. Taken together with the reference metric/vierbein, then in the case of two spin-2 fields, there are 3 different metrics/vierbeins out of which to construct the Lagrangian. The two types of theories considered are referred to as `{\it line theories}' where interactions occur only between pairs of metrics/vierbeins and `{\it cycle theories}' where more than 2 metrics/vierbeins interact in a single vertex (see \cite{Hinterbichler:2012cn,Scargill:2014wya}). These multiple spin-2 theories are modelled on the successful highest cutoff interacting theory of a single spin-2 field, {\it i.e.} ghost-free massive gravity \cite{deRham:2010kj}. \\

In the present work we shall be concerned with the possibility to UV complete the EFTs considered in  \cite{Alberte:2019lnd,Hinterbichler:2012cn}. One of the consistency checks often imposed for low energy EFTs is the requirement for them to have a standard (Lorentz invariant, local and causal) UV completion. This requirement can be imposed by applying the axioms of the S-matrix theory to the scattering amplitudes computed using the low energy EFT, giving constraints on the coefficients in the effective action. These constraints are called `positivity bounds'. There is no proof that the UV completion of theories of massive spin-2 states should satisfy these requirements, and the assumption of locality (exponential/polynomial boundedness) of scattering amplitudes is arguably the weakest requirement. However it is known that by virtue of historical construction, weakly coupled string scattering amplitudes do satisfy these requirements\footnote{Strictly speaking string amplitudes are not polynomially bounded in all directions of complex Mandelstam plane, but they are expected to be so at fixed $t$ \cite{Giddings:2009gj}.}.
The simplest form of these bounds, where only the forward limit of the amplitude is considered, was first derived in \cite{Pham:1985cr,Ananthanarayan:1994hf,Adams:2006sv} and later used to constrain the parameter space of massive spin-2 effective field theories \cite{Cheung:2016yqr,deRham:2017xox,Bonifacio:2016wcb,Bellazzini:2017fep} and to higher derivative corrections of GR \cite{Cheung:2016wjt,Bellazzini:2015cra}. The extension of the positivity bounds beyond forward limit was done in \cite{deRham:2017avq,deRham:2017zjm} and applied to various effective field theories \cite{deRham:2017imi,deRham:2017xox,deRham:2018qqo}. More recently the forward limit positivity bounds have been applied to EFT corrections to the Standard Model \cite{Zhang:2018shp,Bi:2019phv}.
\\

As we shall see, positivity bounds provide remarkably tight constraints on the leading non-derivative interactions in the EFT. Indeed we shall find that similar to the case of a single massless spin-2 field discussed in \cite{Cheung:2016yqr,deRham:2017xox} there is in general a finite region of the multi-dimensional parameter space which is allowed. The reason these constraints are so tight is the fact that even the leading non-derivative interactions are actually irrelevant operators, as is made transparent by means of the \stu formalism described in \cite{Alberte:2019lnd}. Positivity bounds in general do not say anything for renormalizable interactions, but they impose constraints on the signs of combinations of non-renormalizable operators. At the same time, in the case of spinning particles, we may chose arbitrary superpositions of polarization states (indefinite helicity states). By allowing for different indefinite polarizations we can infer distinct inequalities where a given EFT parameter may enter with both positive and negative signs. In this way we can often infer both upper and lower bounds in individual parameters, or more generally compact regions of allowed solutions. At a practical level, we shall find that in general the constraints that arise from arbitrary indefinite polarization scattering can be effectively reproduced by a finite number of well chose indefinite helicity polarization states which we specify in Appendix~\ref{boundsdef}. \\

One might expect that the introduction of additional spin-2 states will increase the allowed parameter region relative to that of a single spin-2 field since there are more free parameters in the EFT Lagrangian. Somewhat surprisingly however we shall find that in general when we include interactions with a second field, they will typically reduce the allowed parameter space region even for the self-interactions compared to what happens with a single field. In other words EFTs with multiple fields may be even more constrained than the individual fields themselves. One reason for this is that the assumed existence of a second spin-2 state gives us increased knowledge of the UV completion. Thus in deriving the positivity bounds, we may choose to subtract the pole contribution of this additional state. If the mass of the second state is comparable to that of the first, we get a much stronger bound than implied by positivity bounds for a single spin-2 state where we would have only subtracted one physical $s$-channel pole (and its $u$-channel conjugate). Indeed with each new field we add to the EFT, we increase the number of poles that may be subtracted, and hence increase the power of the positivity bounds. \\

The remainder of this paper is organized as follows. In section~\ref{sec:main} we briefly introduce the theories under consideration and refer to \cite{Alberte:2019lnd} for a more detailed exposition. In section~\ref{pos bounds} we review the derivation and statement of positivity bounds necessary for the subsequent analysis. We present the results obtained from imposing forward limit positivity bound analysis of various $2$--$2$ scattering processes for the theory with the cycle interactions in section~\ref{sec:cycle}. We discuss the most relevant results in its various subsections. In particular, we find the corrections to the known allowed massive gravity parameter island due to the exchange of the other spin-2 field in subsection~\ref{sec:cycle}\ref{sec:hhhh}.  In section~\ref{2metricssupp} we investigate in detail the theory with suppressed cubic and quartic interactions leading to the highest $\Lambda_3=(m^2 M)^{1/3}$ strong coupling scale. The positivity bounds for the theory with line interactions between the two spin-2 fields are derived in section~\ref{sec:line}. We conclude in section~\ref{sec:conclusions}. We provide the form of the interactions in terms of the mass eigenstates for the line theory in Appendix~\ref{sec:diag}. Our conventions are given in Appendix~\ref{sec:conventions}, while the exact expressions for the scattering amplitudes for cycle theories and the special choice of polarization states are given in Appendix~\ref{boundsdef}.

\section{Cycle and Line EFTs of Interacting Spin-2 Fields}\label{sec:main}

In this work we are interested in the EFTs of two interacting massive spin-2 fields, $h_{\mu\nu}$, $f_{\mu\nu}$. Our goal is to consider EFT interactions with the highest possible cutoff and as such the self-interactions of these states will be based on the $\Lambda_3$ theory of a single massive spin-2 field \cite{deRham:2010ik,deRham:2010kj}, namely ghost-free massive gravity. At the classical level, these ghost-free theories have been extended to multiple massive spin-2 theories in \cite{Hinterbichler:2012cn}. In the EFT context however, what is important is not the full absence of ghosts, but rather the cutoff of the EFT. Demanding only that the EFT is consistent below a given scale allows us to consider a larger class of interactions. To identify the cutoff scale it is most useful to perform a decoupling limit analysis, as was done in detail in \cite{Alberte:2019lnd} for the spin-2 case, based on earlier analyses \cite{ArkaniHamed:2002sp,Creminelli:2005qk,Deffayet:2005ys,deRham:2010ik,Fasiello:2013woa,Ondo:2013wka,Noller:2015eda,Scargill:2015wxs,deRham:2013awa}. \\

In \cite{Alberte:2019lnd} we distinguished between two ways of coupling the two spin-2 fields with the difference arising only at non-linear level. The underlying diffeomorphism symmetry means that it is most convenient to describe the two spin-2 fields in terms of `would-be' metrics $g^{(1)}_{\mu\nu}= (\eta_{\mu\nu}+h_{\mu\nu}/{M_1})^2$, and $g^{(2)}_{\mu\nu} = (\eta_{\mu\nu}+{f_{\mu\nu}}/{M_2})^2$ or equivalently as symmetric vierbeins $e^{(1)}_{\mu\nu}=\eta_{\mu\nu}+h_{\mu\nu}/{M_1}$, $e^{(2)}_{\mu\nu}=\eta_{\mu\nu}+f_{\mu\nu}/{M_2}$. In both cases we supplement each of the fields with a standard Einstein-Hilbert kinetic term, written in terms of these metric/vierbeins, and add non-derivative interactions (\emph{i.e.} a potential) that couple the two fields. In the first class of theories we allow interaction vertices in the potential that couple all three `metrics' $\eta_{\mu\nu}$, $ g^{(1)}_{\mu\nu}$, $g^{(2)}_{\mu\nu}$ together. The quadratic Lagrangian describing this situation is simply a sum of two decoupled Fierz--Pauli theories \cite{Fierz:1939ix}
\ba
\label{eq:2FP}
\L_{\rm FP}=&-& h^{\mu \nu }\mathcal{E}^{\alpha\beta}\mn h\ab-\frac 12 m_1^2 \([h^2]-[h]^2\)\\
&-& f^{\mu \nu }\mathcal{E}^{\alpha\beta}\mn f\ab-\frac 12 m_2^2 \([f^2]-[f]^2\)\,,\nn
\ea
where the standard \Lic operator $\mathcal{E}$ is defined as
\be\label{quadratic}
\mathcal E^{\alpha\beta}_{\mu\nu}h_{\alpha\beta}=-\frac{1}{2}\left[\Box h_{\mu\nu}-\partial_\alpha\partial_{\mu}h^\alpha_{\nu}-\partial_\alpha\partial_\nu h^\alpha_\mu+\partial_\mu\partial_\nu h-\eta_{\mu\nu}\left(\Box h-\partial_\alpha\partial_\beta h^{\alpha\beta}\right)\right]\,,
\ee
and the indices are moved with the Minkowski metric. Here $m_1, m_2$ are the physical masses of $h_{\mu\nu}$ and $f_{\mu\nu}$ respectively. The two spin-2 fields can then be coupled non-linearly through a generic non-derivative interaction $h^n f^m$, with $n+m>2$. We refer to such theories as the `cycle' theories. Note that in principle we could also allow for kinetic/mass mixing at quadratic level but doing so would lower the cutoff as shown in \cite{Alberte:2019lnd}. Nonlinearly the cycle theories thus naturally describe what happens if we take two separate copies of nonlinear ghost-free massive gravity, and couple them together with interactions at the highest possible scale. \\

The second class of theories considered below is what we refer to as `line' theories. In this case there is mass mixing between the two fields $\tilde h_{\mu\nu}$, $\tilde f_{\mu\nu}$ already at linear level and the two remain coupled also non-linearly. Moreover, the non-derivative interaction terms are such that only one of the metrics ($g^{(1)}_{\mu\nu}= (\eta_{\mu\nu}+\tilde h_{\mu\nu}/{M_1})^2$ in our case) couples to the flat reference metric $\eta_{\mu\nu}$. The second metric, $g^{(2)}_{\mu\nu}=(\eta_{\mu\nu}+\tilde f_{\mu\nu}/{M_1})^2$, in turn has direct couplings only to $g^{(1)}_{\mu\nu}$. The quadratic Lagrangian for the line of interactions is given by
\begin{align}
\label{eq:line}
\L_{\rm line}=&- \tilde h^{\mu \nu }\mathcal{E}^{\alpha\beta}\mn \tilde h\ab-\frac 12 \tilde m_1^2 \([\tilde h^2]-[\tilde h]^2\)\\
&- \tilde f^{\mu \nu }\mathcal{E}^{\alpha\beta}\mn \tilde f\ab-\frac 14 \tilde m_2^2 \([(\tilde f-\tilde h)^2]-[\tilde f-\tilde h]^2\)\,.\nn
\end{align}
While on the first line we see the usual Fierz--Pauli mass term for $\tilde h_{\mu\nu}$ written with respect to $\eta_{\mu\nu}$ the mass term for $\tilde f_{\mu\nu}$ is written with respect to $\tilde h_{\mu\nu}$.
Importantly, with the action given in this form, the fields $\tilde h_{\mu\nu}$, $\tilde f_{\mu\nu}$ are not the mass eigenstates and the parameters $\tilde m_1, \tilde m_2$ do not represent the physical masses --- the action first needs to be diagonalized. The diagonalization can be done by a simple rotation in the field space leading to an action of the exactly same form as \eqref{eq:2FP}. However, once \eqref{eq:line} is supplemented with the non-linear interactions --- both the non-linear Einstein--Hilbert terms and the non-derivative couplings --- this procedure leads to a non-trivial kinetic mixing between the two diagonalized fields. We shall present the full details of how this mixing occurs below in subsection \ref{sec:line_intro}. \\

The lowest derivative interactions in a generic non-linear EFT for two interacting massive spin-2 fields with either cycle or line interactions is most conveniently written in terms of the `would-be' metrics $g^{(1)}$ and $g^{(2)}$ as
\ba
\label{genericeft}
 \hspace{-10pt} g_*^2 \L_{\rm EFT}[g^{(1)},g^{(2)}]=\sum_{i=1,2}\sqrt{-g^{(i)}}\frac{M_i^2}{2}R[g^{(i)}]+\frac{m^2M^2}{4}\sqrt{-g^{(1)}}\mathcal L_{\rm int}[\eta, g^{(1)},g^{(2)}]\,,
 \ea
 where $M_i$ represents the scale of non-linearities contained in the Einstein--Hilbert kinetic terms and $m\ll M_i$, $M\sim M_i$ are mass scales that we shall specify below. The effective Planck scale associated to the two spin-2 fields is then related to the weak coupling parameter $g_*$ roughly\footnote{In the case of line interactions the exact physical Planck scales can only be determined after the diagonalization but they are always given parametrically by $M_1$ and $M_2$.} as
 \be
M_{\text{Pl}} \sim \frac{M}{g_*} \, .
\ee
It is required by the so-called improved positivity bounds  \cite{Bellazzini:2016xrt,deRham:2017imi,deRham:2017xox} that $g_* \ll 1$. For an EFT of two interacting spin-2 fields with the strong coupling scale $\Lambda_3=(m^2M)^{1/3}$ this imposes \cite{deRham:2017xox,Bellazzini:2017fep}
\be
g_* \lesssim \frac{m}{\Lambda_3}\ll 1 \, .
\ee
Provided this assumption is made, then we may focus on tree level positivity bounds in what follows since the smallness of $g_*$ allows us to ignore loop corrections. In the following subsections we present the explicit form of \eqref{genericeft} for the cycle and line theories.

\subsection{Cycle of Interactions}\label{sec:cycle_intro}
Within the framework of cycles of interactions, the non-linear action describing the fields $h_{\mu\nu}$ and $f_{\mu\nu}$ can be written as \cite{Alberte:2019lnd}
\be\label{action}
\begin{split}
g_*^2 \L_{\rm cycle}=&\frac{M_1^2}{2}\sqrt{-g^{(1)}}\,R[g^{(1)}]+\frac{m_1^2M_1^2}{4}\sqrt{-\eta}\,\sum_{n=0}^4\kappa^{(1)}_n\,\U_n\left[\eta^{-1}h/M_1\right]\\
+&\frac{M_2^2}{2}\sqrt{-g^{(2)}}\,R[g^{(2)}]+\frac{m_2^2M_2^2}{4}\sqrt{-\eta}\,\sum_{n=0}^4\kappa^{(2)}_n\,\U_n\left[\eta^{-1}f/M_2\right]\\
+&\frac{m^2M_1M_2}{4}\L_{\text{int}}[h/M_1,f/M_2] +\L_{\text{h.d.}}\,,
\end{split}
\ee
where the spin-2 fields are defined as $g^{(1)}_{\mu\nu}= (\eta_{\mu\nu}+h_{\mu\nu}/{M_1})^2$, and $g^{(2)}_{\mu\nu} = (\eta_{\mu\nu}+{f_{\mu\nu}}/{M_2})^2$ and where $\L_{\text{h.d.}}$ denotes higher derivative terms that arise in the effective theory discussed in more detail in \cite{Alberte:2019lnd}.
The potential terms $\mathcal U_n$ are of the double-epsilon structure and for any matrix $\mathbb X$ can be expressed in terms of the flat space Levi-Civita tensor as
\begin{equation}\label{mass1}
\U_n(\mathbb X)=\varepsilon_{\mu_1\dots\mu_n\mu_{n+1}\dots\mu_4}\varepsilon^{\nu_1\dots\nu_n\nu_{n+1}\dots\nu_4}\mathbb X^{\mu_1}_{\nu_1}\dots\mathbb X^{\mu_n}_{\nu_n}\delta^{\mu_{n+1}}_{\nu_{n+1}}\dots\delta^{\mu_4}_{\nu_4}\equiv\varepsilon\varepsilon I^{4-n}\mathbb X^n\,,
\end{equation}
where in the last equality we have introduced shorthand notations as in \cite{Ondo:2013wka} that will be used throughout this work. Then for instance we have
\begin{equation}\label{mass1}
\U_3(\eta^{-1}h)=\varepsilon_{\mu\nu\alpha\beta}\varepsilon^{\mu'\nu'\alpha'\beta'}\delta^\mu_{\mu'} h^{\nu}_{\nu'} h^{\alpha}_{\alpha'} h^{\beta}_{\beta'}\equiv\varepsilon\varepsilon Ihhh\,,\quad\text{etc.}
\end{equation}
Henceforth we set $\kappa^{(i)}_0=\kappa^{(i)}_1=0$ in order to impose the no tadpole condition and fix the constant term to zero. We also set $\kappa^{(i)}_2=1$ in order to normalize the masses of the fields to $m_i$ while $\kappa^{(i)}_3$ and $\kappa^{(i)}_4$ remain free parameters.

The interaction term, $\mathcal L_{\rm int}$, is again a sum of double-epsilon interactions between the two fields parameterized as:
\be\label{def_int}
\L_{\text{int}}[h,f]=2c_1\L_{hhf}+2c_2\L_{hff}+\lambda\L_{hhff} + d_1 \L_{hhhf}+ d_2 \L_{hfff}\,,
\ee
with $c_1,c_2, \lambda,d_1, d_2$ --- arbitrary dimensionless coefficients. The various interaction terms are given by
\begin{equation}\label{def_int_2}
\L_{hhf}=\varepsilon\varepsilon I hhf, \ \L_{hff}=\varepsilon\varepsilon I h ff, \
\L_{hhff}=\varepsilon\varepsilon hhff,\ \L_{hhhf}=\varepsilon\varepsilon hhhf,\  \L_{hfff}=\varepsilon\varepsilon h f f f.
\end{equation}
The last two interactions ($d_1, d_2$) have been included for completeness but they do not contribute to any elastic $2-2$ scattering (at tree-level) and are therefore blind to current positivity bounds.

We also fix the mass scalings as
\be\label{scaling0}
m_2\equiv m\,,\quad \frac{m_1}{m_2}\equiv x\,;\qquad M_2\equiv\M\,,\quad\frac{M_1}{M_2}\equiv\gamma\,.
\ee
Whenever discussing theories with cycle interactions we will focus on the case where there is no large separation of scales between $m_1$ and $m_2$ and between $M_1$ and $M_2$. In particular when looking at specific configurations we will always limit ourselves to $10^{-1}<\gamma<10$.
We also restrict ourselves to the region where the mass ratio is $1/2<x<2$, (\emph{i.e.} $ m_{1}<2m_{2}<4m_1$), so as to avoid decay of the heavier field into the lighter one, \cite{deRham:2017zjm}.

The action \eqref{action} above can be rewritten in a more standard form, used in the context of a single massive spin-2 field, as \cite{deRham:2010kj,deRham:2014zqa}
\begin{equation}\label{act0}
g_*^2 \L_{\rm cycle}=\sum_{i=1}^2\,\sqrt{-g^{(i)}}\left[\frac{M_i^{2}}{2}R[g^{(i)}]+\frac{m_i^2}{2}\,\sum_{n=0}^4\alpha^{(i)}_n\,\U_n\left[\mathcal K(g^{(i)},\eta)\right]\right]+\frac{m^2M_1M_2}{4}\L_{\text{int}}[\eta, g^{(1)},g^{(2)}] +\L_{\text{h.d.}}\;,
\end{equation}
where the tensor $\mathcal K^\mu_\nu$ is defined as
\ba
\label{eq:K}
\mathcal K\mupn(g^{(i)},\eta)=\delta^\mu_\nu-\left(\sqrt{g^{-1}_{(i)}\eta}\right)\mupn\;,
\ea
and the coefficients $\alpha_3^{(i)}$, $\alpha_4^{(i)}$ are related to the $\kappa_n^{(i)}$'s in \eqref{action} through
\be
\kappa^{(i)}_3 =2+\alpha^{(i)}_3 \,,\quad\kappa^{(i)}_4 = 1+\alpha^{(i)}_3+\alpha^{(i)}_4\,.
\ee
This form of the action will be useful when comparing the results obtained from the positivity bounds in section~\ref{sec:cycle} with previous results for a single spin-2 field \cite{Cheung:2016yqr}. \\

For future reference we write the full Lagrangian that will be used for the positivity bounds analysis explicitly (ignoring the $d_1$ and $d_2$ mixed interactions and higher derivative terms):
\be\label{model}
\begin{split}
g_*^2 \L_{\rm cycle}=&\gamma^2\frac{\M^2}{2}\sqrt{-g^{(1)}}R[g^{(1)}]+x^2\frac{m^2}{4}\left[\varepsilon\varepsilon I^2hh+\frac{\kappa_3^{(1)}}{\gamma M}\varepsilon\varepsilon Ihhh+\frac{\kappa_4^{(1)}}{\gamma^2 M^2}\varepsilon\varepsilon hhhh\right]\\
+&\frac{\M^2}{2}\sqrt{-g^{(2)}}R[g^{(2)}]+\frac{m^2 }{4}\left[\varepsilon\varepsilon I^2ff+\frac{\kappa_3^{(2)}}{M}\varepsilon\varepsilon Ifff+\frac{\kappa_4^{(2)}}{M^2}\varepsilon\varepsilon ffff\right]\;\\
&+\frac{m^2}{4 \gamma M}\left[2c_1\varepsilon\varepsilon Ihhf+2c_2 \gamma \varepsilon\varepsilon Iffh+\frac{\lambda}{M}\varepsilon\varepsilon hhff\right]\,.
\end{split}
\ee
It was established in \cite{Alberte:2019lnd} that this theory becomes strongly coupled at the scale $\Lambda_{7/2}=(m^{5/2}M)^{2/7}$. This happens due to interactions in the helicity-0/helicity-1 sector of the theory and can only be avoided in the case of special (technically natural) tuning of the coupling constants $\{c_{1,2}, \lambda, d_{1,2}\}\to m/\Lambda_3\cdot\{c_{1,2}, \lambda, d_{1,2}\} $ leading to a $\Lambda_3$ theory. We shall present the constraints coming from the positivity bounds on the forward $2-2$ scattering amplitudes in section \ref{sec:cycle} for the general case and in section \ref{2metricssupp} for the rescaled $\Lambda_3$ case.

\subsection{Line of Interactions}\label{sec:line_intro}
As explained earlier in this section, an alternative way of describing the dynamics of two interacting massive spin-2 fields is to consider a theory with a line of interactions. In this case, the two fields, $\tilde h_{\mu\nu}$ and $\tilde f_{\mu\nu}$, are mixed already at linear level as in \eqref{eq:line}. The corresponding non-linear theory is then obtained by first building the tensor $\K\mupn$ out of the `would-be' metrics $g\mn^{(1,2)}$ as
\ba\label{Kmunu}
\K\mupn(g^{(1)},g^{(2)})= \delta \mupn -\(\sqrt{g_{(1)}^{-1}g_{(2)}}\)\mupn\,
\ea
and working with the same conventions
\be\label{metric_def}
\begin{split}
g^{(1)}\mn&=\(\eta\mn+\tilde h\mn/M_1\)^2\,,\qquad g^{(2)}\mn=\(\eta\mn+\tilde f\mn/M_2\)^2\,.
\end{split}
\ee
We should note that in \cite{Alberte:2019lnd} in order to derive the decoupling limit it was more helpful to use a different decomposition for the second metric, namely $g_{\mu\nu}^{(2)}\equiv(g_{\mu\alpha}^{(1)}+\tilde f_{\mu\alpha})g^{\alpha\beta}_{(1)}(g_{\beta\nu}^{(1)}+\tilde f_{\beta\nu})$. In the context of computing scattering amplitudes, the choice of field variables does not matter, and so for ease of comparison with the cycle case we will use the same convention \eqref{metric_def}.

The interactions between $g^{(1)}$ and $g^{(2)}$ that have the highest possible cutoff are the double-epsilon polynomials of  $\K$ \cite{Alberte:2019lnd}. The action corresponding to a line of interactions with the highest possible cutoff is thus given by
\ba\label{eq:actionline}
g_*^2 \L_{\rm line}&=&\frac{M_1^2}{2}\sqrt{-g^{(1)}}\,R[g^{(1)}]
+\frac{\tilde m_1^2M_1^2}{4}\sqrt{-\eta}\,\sum_{n=2}^4\tilde \alpha_n\,\U_n\left[\eta^{-1}\tilde h/M_1\right]\\
&+&\frac{M_2^2}{2}\sqrt{-g^{(2)}}\,R[g^{(2)}]+\frac{\tilde m_2^2M^2}{4}\sqrt{-g^{(1)}}\,\sum_{n=2}^4\tilde \beta_n\,\U_n\left[\K(g^{(1)},g^{(2)})\right]
+\L_{\text{h.d.}}\,,\nn
\ea
where the sum starts from $n=2$ as needed to ensure the absence of tadpoles and to set the constant term to zero. We use again the convention $\tilde \alpha_2=\tilde \beta_2=1$ and define
\ba
M^2\equiv  \frac{M^2_1 M^2_2}{M^2_1+M^2_2}\,.
\ea
Similarly to the case of the cycle interactions we introduce the mass scalings
\be\label{scaling2}
\tilde m_2\equiv  m\,,\quad \frac{\tilde m_1}{\tilde m_2}\equiv \tilde x\,;\quad \frac{M_1}{M_2}\equiv\gamma\,.
\ee
In these notations $M^2/M_2^2=\gamma^2/(1+\gamma^2)$ so that $M^2\to M_2^2$ when $\gamma\to \infty$. At linear order in fields one can expand $\mathcal K$ as
\be
\K\mupn(g^{(1)},g^{(2)})=\frac{\tilde h\mupn}{M_1}-\frac{\tilde f\mupn}{M_2} + \dots,
\ee
and thus the action \eqref{eq:actionline} takes a form similar to that of  \eqref{eq:line}.
As before, this introduces a mixing between the two fields $\tilde h_{\mu\nu}$, $\tilde f_{\mu\nu}$ present already in the quadratic mass terms. In other words, the fields $\tilde h_{\mu\nu}, \tilde f_{\mu\nu}$ are not the mass eigenstates. As a result, the masses $\tilde m_1, \tilde m_2$ are not the physical masses and the scales $M_1,M_2$ are not the physical coupling scales. It is also important to emphasize that $\gamma$ defined in \eqref{scaling2} and $\gamma$ defined in \eqref{scaling0} are different. While the latter really represents the ratio between the two physical scales of non-linearities, the former although similar is a formal ratio between the two scales $M_{1,2}$ appearing in the action \eqref{eq:actionline}.

The diagonalization that needs to be performed in the action \eqref{eq:actionline} can be carried out by a rotation in field space given by
\ba
\tilde h\mn &=& \cos \theta h\mn+\sin \theta f\mn\\
\tilde f\mn &=& -\sin \theta h\mn+\cos \theta f\mn\,,
\ea
with rotation angle satisfying
\ba\label{rot_angle}
\tan 2 \theta = \frac{2 M_1 M_2 \tilde m_2^2}{M_1^2 (\tilde m_1^2-\tilde m_2^2)+M_2^2 (\tilde m_1^2+\tilde m_2^2)}\,.
\ea
The resulting diagonalized quadratic Lagrangian is then
\ba
\label{eq:line2diag}
\L^{(2)}_{\rm line}=- h^{\mu \nu }\mathcal{E}^{\alpha\beta}\mn  h\ab-   f^{\mu \nu }\mathcal{E}^{\alpha\beta}\mn  f\ab-\frac 12  m_1^2 \([ h^2]-[ h]^2\)-\frac 12  m_2^2 \([ f^2]-[f]^2\)\,,
\ea
with the physical masses $m_{1,2}$ given by
\ba\label{masses}
m_{1,2}^2=\frac12\(
\tilde m_1^2+\tilde m_2^2\pm \frac{M_1^2 (\tilde m_1^2-\tilde m_2^2)^2+M_2^2 (\tilde m_1^2+\tilde m_2^2)^2}
{M_1^2 (\tilde m_1^2-\tilde m_2^2)^{\phantom{2}}+M_2^2 (\tilde m_1^2+\tilde m_2^2)^{\phantom{2}}}\cos 2\theta
\)\,.
\ea
This introduces a further non-linear mixing in the kinetic terms: the cubic and quartic kinetic terms are symbolically of the form
\ba\label{cubic_kinetic}
\L^{(3)}_{\rm Kin}&=&\frac{1}{M_1}\p^2 (\C h + \S f)^3+\frac{1}{M_2}\p^2 (\C f - \S h)^3\,,\\\label{quartic_kinetic}
\L^{(4)}_{\rm Kin}&=&\frac{1}{M_1^2}\p^2 (\C h + \S f)^4+\frac{1}{M_2^2}\p^2 (\C f - \S h)^4\,,
\ea
where $\p^2 h^3 $, (resp.  $\p^2 h^4$) is the symbolic representation of the standard Einstein-Hilbert term at cubic order (resp. quartic order), in the convention where the metrics are given in \eqref{metric_def}.

As for the potential terms, they take the standard form (as in the case of cycle interactions) with redefined coefficients with an additional quartic mixing,
\ba\label{line_cubic}
\L^{(3)}_{\rm mass}&=&\frac{\tilde m_2^2}{4 M_1}\ \sum_{n=0}^3
\kappa^{(3)}_{n} \varepsilon\varepsilon I h^{3-n} f^{n} \,,\\\label{line_quartic}
\L^{(4)}_{\rm mass}&=&\frac{\tilde m_2^2}{4 M_1^2}\ \sum_{n=0}^4
\kappa^{(4)}_{n} \varepsilon\varepsilon h^{4-n} f^{n}+\frac{ \tilde m_2^2}{ 4(M_1^2+ M_2^2)}\([f\cdot h\cdot f\cdot h]-[f^2\cdot h^2]\)\,.
\ea
The expressions for the coefficients $\kappa^{(3,4)}_n$ are given in Appendix~\ref{sec:diag}. Even though we have introduced nine different parameters $\kappa^{(3)}_{0,\cdots,3}$ and $\kappa^{(4)}_{0,\cdots,4}$, the theory only has four independent coupling constants $\tilde \alpha_{3,4}$ and $\tilde \beta_{3,4}$ in addition to the ratio $M_1/ M_2\equiv \gamma$ and the angle $\theta$ (roughly representing the ratio between the physical masses $m_1$ and $m_2$). The latter can be expressed in terms of $\tilde x$ and $\gamma$ defined in \eqref{scaling2}, so that the full set of independent couplings here are $\{\tilde\alpha_3,\tilde\alpha_4,\tilde\beta_3,\tilde\beta_4,\tilde x, \gamma\}$. The various mass and interaction terms appearing in \eqref{eq:line2diag}, \eqref{line_cubic}, \eqref{line_quartic} are almost identical in structure to the terms appearing for the cycle of interactions in \eqref{model}. We note however that the coefficients in front of the various terms are different (in particular, the $\kappa^{(i)}_n$'s appearing in \eqref{model} are not the $\kappa^{(3,4)}_n$'s appearing in \eqref{line_cubic} and \eqref{line_quartic}). We give the list of all couplings controlling the various interactions for both cycle and line interactions in Table~\ref{tab:compare}.

The final Lagrangian (up to quartic order in fields $f_{\mu\nu}$, $h_{\mu\nu}$) for a theory with line interactions that will be used for the positivity bounds analysis in section~\ref{sec:line} is the sum of all the terms given above:
\be\label{action_line}
g_*^2 \L_{\rm line}=\L^{(2)}_{\rm line}+\L^{(3)}_{\rm Kin}+\L^{(4)}_{\rm Kin}+\L^{(3)}_{\rm mass}+\L^{(4)}_{\rm mass}\,.
\ee
From the extended analysis of \cite{Alberte:2019lnd} it is known that this theory becomes strongly coupled at $\Lambda_3$ and is thus indeed the line theory of two interacting massive spin-2 fields with interactions with the highest possible interaction scale.

\begin{table}[h]\small
    \centering
    \begin{tabular}{|c|c|c|}
   \hline
         Interaction&Cycle & Line   \\
         \hhline{|=|=|=|}
                  $\varepsilon\varepsilon I^2h^2$ &$\frac{m^2}{4}x^2$ & $\frac{m_2^2}{4}x^2$  $\phantom{\Bigg]}$\\[0.2cm]
         \hline
         $\varepsilon\varepsilon I^2f^2$ &$\frac{m^2}{4}$& $\frac{m_2^2}{4}$ $\phantom{\Bigg]}$\\[0.2cm]
         \hhline{|=|=|=|}
$\varepsilon\varepsilon Ih^3$&$\frac{m^2x^2}{4\gamma M}\kappa_3^{(1)}$&$\frac{m^2}{4\gamma M_2}\kappa_0^{(3)}$ $\phantom{\Bigg]}$\\[0.2cm]
         \hline
         $\varepsilon\varepsilon If^3$ &$\frac{m^2x^2}{4M}\kappa^{(2)}_3$& $\frac{m^2}{4\gamma M_2}\kappa^{(3)}_3$    $\phantom{\Bigg]}$\\[0.2cm]
    \hline
    $\varepsilon\varepsilon I h^2f$&$\frac{m^2}{4\gamma M}2c_1$&$\frac{m^2}{4\gamma M_2}\kappa^{(3)}_1$ $\phantom{\Bigg]}$\\[0.2cm]
    \hline
    $\varepsilon\varepsilon I hf^2$ &$\frac{m^2}{4M}2c_2$&$\frac{m^2}{4\gamma M_2}\kappa_2^{(3)}$ $\phantom{\Bigg]}$\\[0.2cm]
  \hhline{|=|=|=|}
  $\varepsilon\varepsilon h^4$&$\frac{m^2x^2}{4\gamma^2 M^2}\kappa^{(1)}_4$&$\frac{m^2}{4\gamma^2M_2^2}\kappa^{(4)}_0$ $\phantom{\Bigg]}$\\[0.2cm]
  \hline
  $\varepsilon\varepsilon f^4$&$\frac{m^2}{4M^2}\kappa_4^{(2)}$&$\frac{m^2}{4\gamma^2M_2^2}\kappa^{(4)}_4$ $\phantom{\Bigg]}$\\[0.2cm]
  \hline
  $\varepsilon\varepsilon h^2 f^2$&$\frac{m^2}{4\gamma M^2}\lambda$&$\frac{m^2}{4\gamma^2M_2^2}\kappa^{(4)}_2$ $\phantom{\Bigg]}$\\[0.2cm]
  \hline
  $\varepsilon\varepsilon h^3 f$&$\frac{m^2}{4\gamma^2M^2}d_1$&$\frac{m^2}{4\gamma^2M_2^2}\kappa^{(4)}_1$ $\phantom{\Bigg]}$\\[0.2cm]
  \hline
  $\varepsilon\varepsilon hf^3$&$\frac{m^2}{4M^2}d_2$&$\frac{m^2}{4\gamma^2M_2^2}\kappa_2^{(4)}$ $\phantom{\Bigg]}$\\[0.2cm]
  \hline
  $[f\cdot h\cdot f\cdot h]-[f^2\cdot h^2]$&$0$&$\frac{ \tilde m_2^2}{ 4(M_1^2+ M_2^2)}$ $\phantom{\Bigg]}$\\[0.2cm]
  \hline
    \end{tabular}
    \caption{Comparison between the coupling constants appearing in front of the mass and interaction terms in the action for cycle and line interactions, \eqref{model} and \eqref{action_line} respectively. }
    \label{tab:compare}
\end{table}

\section{Positivity Bounds}\label{pos bounds}

With both the cycle and line EFTs for two interacting spin-2 fields with the highest possible interaction scale ($\Lambda_{7/2}$ or $\Lambda_3$ respectively) at hand, we now turn to the main question addressed in this work of whether these EFTs could in principle admit a healthy UV completion.  In general this is not a question that can be answered without assumptions about the precise nature of that UV completion, and the properties demanded of it. The most conservative assumptions are those borrowed from the 1960's S-matrix program and axiomatic field theory approaches. They assume that the UV completion is Lorentz invariant, local, causal and unitary. Causality is implemented in the S-matrix by demanding analyticity (modulo physical poles and branch cuts) in slices of the complex Mandelstam plane (which may be extended to some form of maximal analyticity). Locality imposes the requirement that the scattering amplitudes are polynomially (strictly exponentially) bounded which allows us to write dispersion relations with a finite number of subtractions.  \\

These assumptions on the UV theory were well founded in the context of the historical S-matrix program where the majority of interest was in non-gravitational theories with a mass gap. The existence of a mass gap implies a finite region of analyticity in the complex Mandelstam plane for scattering amplitudes where amplitudes related by crossing symmetry are seen to be identical. In addition polynomial boundedness and a mass gap implies the Froissart-Martin bound \cite{Froissart:1961ux,Martin:1965jj,PhysRev.135.B1375} for the growth of the $2-2$ scattering amplitudes. Taken together, this allows us to write dispersion relations with finite numbers of subtractions for amplitudes of all spin (the dispersion relations for general spin with desired positivity and crossing symmetry properties were recently given in \cite{deRham:2017zjm}). The validity of these assumptions in the context of gravitational theories, such as theories of quantum gravity like string theory are more questionable. In particular, the assumption of locality in a gravitational theory is unclear when the metric defining the lightcone is fluctuating \cite{deRham:2019ctd,Giddings:2009gj,Keltner:2015xda}. \\

In the context of the current EFTs, we do have a mass gap, and so if we make the assumption that the UV completion has polynomially bounded scattering amplitudes, then we may safely assume the spin-2 version of the Froissart-Martin bound. In this case we can use the properties of S-matrix analyticity, unitarity and crossing symmetry to put constraints on combinations of coefficients of operators in the above constructed Wilsonian effective action for the cycle and line theories, following the approach of~\cite{Pham:1985cr,Ananthanarayan:1994hf,Adams:2006sv}. These constraints, also known as \emph{positivity bounds}, can be applied both in the forward scattering limit, and for general spins away from the forward scattering limit \cite{deRham:2017zjm,deRham:2018qqo}. This has been a powerful consistency check for effective field theories admitting a Lorentz invariant UV completion.\\

In the following we shall apply these positivity bounds in the forward limit, $t=0$,  for indefinite elastic scattering of two massive spin-2 particles, $m_1$ and $m_2$ (Appendix~\ref{sec:conventions} contains a summary of all our conventions). Our reasons for focusing on the forward limit is that, as discussed in \cite{deRham:2018qqo}, the non-forward limit bounds derived in \cite{deRham:2017zjm} generalizing \cite{deRham:2017avq} are most useful to constraining parameters in the $\Lambda_5$ effective theory, whereas for the $\Lambda_3$ theory they only give a marginal improvement over the (simpler to derive) forward limit bounds for indefinite scattering amplitudes discussed in \cite{Cheung:2016yqr}. In short then, the following analysis will extend the work of  Cheung and Remmen \cite{Cheung:2016yqr} to two interacting massive spin-two fields. Although not our main focus, we note that there have been interesting works constraining EFTs using the requirement of asymptotic (sub)luminality \cite{Camanho:2016opx,Hinterbichler:2017qyt,Bonifacio:2017nnt,Hinterbichler:2017qcl}, at least for weakly coupled UV completions. In brief these require that the Eisenbud-Wigner scattering time-delay \cite{Wigner:1955zz}, which can be inferred from the eikonal scattering limit, is positive. In the case of a single massive spin-2 (and massless spin-2) in \cite{Hinterbichler:2017qyt,Bonifacio:2017nnt} it is argued that these conditions restrict to corresponds to a one-parameter family of the ghost-free massive gravity model. These constraints are distinct from those following from positivity. It would be interesting to apply them to the two spin-2 case considered here. \\

\subsection{Formalism}
\begin{figure}[H]
    \centering
\includegraphics[width=0.95\textwidth]{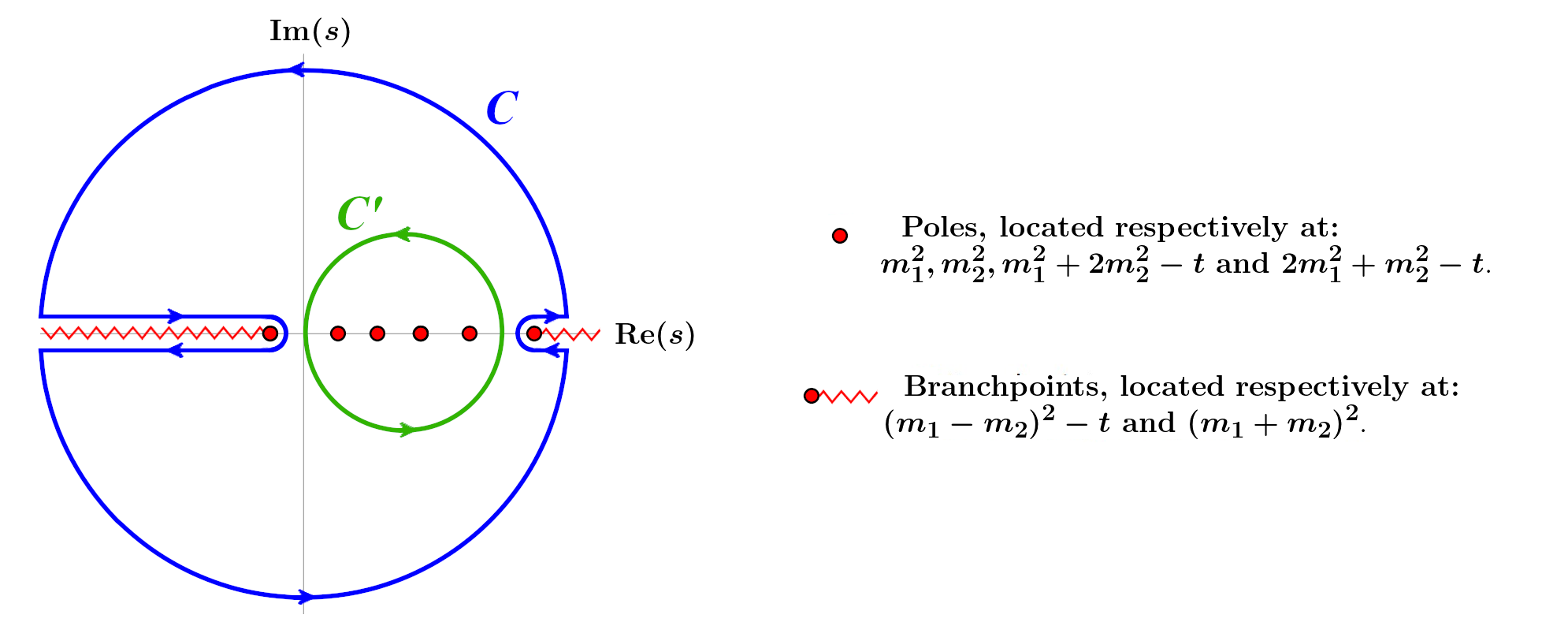}\\
\caption{Analytic structure of the {$2-2$ elastic scattering} amplitude in the complex $s$ plane for spin-zero particles or for regularized combinations of general spin transversity amplitudes \cite{deRham:2017zjm}. We show the analytic structure for arbitrary values of $t$ but calculate bounds in the forward limit, $t=0$. }
\label{fig:defin11}
\end{figure}

 Here we review briefly the derivation of the forward limit positivity bounds for general spin given in  \cite{Bellazzini:2016xrt}. On considering the scattering amplitude for two particles of different masses and spin-zero, in terms of Mandelstam variables $s,t,u$,  $s$ being the center of mass energy squared, the analytic structure is given by Fig.~\ref{fig:defin11}. For general spins away from the forward limit, the same analytic structure holds only for special combinations of regularized transversity scattering amplitudes as discussed in \cite{deRham:2017zjm}. In the forward scattering limit this complication is largely avoided (at least for boson scattering \cite{Bellazzini:2016xrt}). In the complex $s$ plane, the $2-2$ elastic scattering amplitude has four physical poles located at $s = m_1^2$, $s = m_2^2$, $s = 2m_1^2+m_2^2-t$ and $s= m_1^2+2m_2^2-t$ and two branch cuts starting from $s = (m_1+m_2)^2$ to infinity and from $s=(m_1-m_2)^2-t$ to minus infinity. In the remaining $s$ plane the scattering amplitude must be analytic and bounded for fixed $\text{Min}(m_1^2,m_2^2) >t \ge 0$ in the form $\lim_{|s| \rightarrow \infty} |{A}(s,t)| < c \, s^2$ for some constant $c$ \cite{PhysRev.135.B1375}.
 We will for convenience consider  $m_{1}<2m_{2}<4m_1$ or $1/2<x<2$ to prevent the heavier particle from decaying into the lighter one, \cite{deRham:2017zjm}. This also ensures a clean separation between the poles and the branch cuts in the analytic structure of the amplitude in the complex $s$ plane. \\

We then write a twice subtracted dispersion relation for the $s-$channel scattering amplitude in the complex $s$ plane using Cauchy's integral formula where the singularities in the $u$-channel are also included by virtue of crossing symmetry
\begin{equation}
\begin{split}
\hspace{-2cm}A^s_{\lambda_1\lambda_2\lambda_3\lambda_4}(s,0)&=a^s_{\lambda_1\lambda_2\lambda_3\lambda_4}+b^s_{\lambda_1\lambda_2\lambda_3\lambda_4} (s-u)+ \frac{\text{Res} A^s_{\lambda_1\lambda_2\lambda_3\lambda_4}(s=m_{1}^2,0)}{m_{1}^2-s}+\frac{\text{Res} A^s_{\lambda_1\lambda_2\lambda_3\lambda_4}(s=m_{2}^2,0)}{m_{2}^2-s} \hspace{-2cm}\\
&+\frac{\text{Res}A^u_{\lambda_1\bar \lambda_4 \lambda_3 \bar \lambda_2}(u=m_{1}^2,0)}{m_{1}^2-u} +\frac{\text{Res}A^u_{\lambda_1\bar \lambda_4 \lambda_3 \bar \lambda_2}(u=m_{2}^2,0)}{m_{2}^2-u} \\
&+\frac{(s-s_0)^2}{2\pi i }\int_{(m_{1}+m_{2})^{2}}^{\infty}\d s'\left(\frac{\text{Disc}_{s}A^s_{\lambda_1\lambda_2\lambda_3\lambda_4}(s',0)}{(s'-s_0)^2(s'-s)}+\frac{\text{Disc}_{u}A^u_{\lambda_1\bar \lambda_4\lambda_3 \bar \lambda_2}(s',0)}{(s'-s_0)^2(s'-u)}\right) \,.
\end{split}
\end{equation}
Here Res stands for the residue and the discontinuity across the branch cuts is defined as $\text{Disc}_s A(s)=2i \text{Abs}_s A(s)\equiv \lim_{\epsilon\to 0} A(s+i \epsilon)-A(s-i \epsilon)$, $\lambda_1\lambda_2\lambda_3\lambda_4$ denote the helicities or polarization states of the incoming and outgoing particles. $a^s_{\lambda_1\lambda_2\lambda_3\lambda_4}=a^u_{\lambda_1 \bar \lambda_4 \lambda_3 \bar \lambda_2}$ and $b^s_{\lambda_1\lambda_2\lambda_3\lambda_4}=-b^u_{\lambda_1 \bar \lambda_4 \lambda_3 \bar \lambda_2}$ are the subtraction constants defined at some arbitrary subtraction scale $s_0$.
This dispersion relation manifests crossing symmetry in the form
\be
A^s_{\lambda_1\lambda_2\lambda_3\lambda_4}(s,0,u)=A^u_{\lambda_1\bar \lambda_4 \lambda_3 \bar \lambda_2}(u,0,s) \, .
\ee
In the present case, the $s$ and $u$ channel scattering amplitudes are identical since we focus on elastic scattering amplitudes of the form $A+B \rightarrow A+B$, with the only difference being the flip in helicities $\bar \lambda=-\lambda$, and for indefinite polarizations -- complex conjugation of polarization tensors.

By using the optical theorem, which is derived from the unitarity of the S-matrix, and restricting to elastic scattering $\lambda_3=\lambda_1$, $\lambda_4=\lambda_2$, we can show that the integral across the branch cuts is positive since in the physical region the absorptive part of the associated $s$ and $u$ channel reactions are
\ba
&& \text{Abs}_s \,  A^s_{\lambda_1\lambda_2\lambda_1\lambda_2}(s,0) = \sqrt{(s-m_1^2-m_2^2)^2-4m_1^2m_2^2} \, \sigma^{\rm total}(\lambda_1,\lambda_2 \rightarrow \text{total}) >0 \, , \\
&& \text{Abs}_u \, A^u_{\lambda_1 \bar \lambda_2 \lambda_1 \bar \lambda_2}(u,0) = \sqrt{(u-m_1^2-m_2^2)^2-4m_1^2m_2^2} \, \sigma^{\rm total}(\lambda_1,\bar \lambda_2 \rightarrow \text{total}) >0 \, .
\ea
After taking the second derivative of the pole subtracted amplitude, the subtraction constants vanish and hence we have:
\begin{equation}
\begin{split}\label{pos}
      f_{\lambda_1\lambda_2}&=\frac{1}{2}\frac{\d^2}{\d s^2}(A^s_{\lambda_1\lambda_2\lambda_1\lambda_2}(s,0)-\text{poles})\\
      &=\frac{1}{2\pi i}\oint \d s' \frac{(A^s_{\lambda_1\lambda_2\lambda_1\lambda_2}(s,0)-\text{poles})}{(s'-s)^3}\\
      &=\frac{1}{\pi }\int_{(m_{1}+m_{2})^{2}}^{\infty}\d s'\left(\frac{\text{Abs}_s A^s_{\lambda_1\lambda_2\lambda_1\lambda_2}(s',0)}{(s'-s)^3}+\frac{\text{Abs}_u A^u_{\lambda_1 \bar \lambda_2\lambda_1 \bar \lambda_2}(s',0)}{(s'-u)^3}\right)>0\,,
\end{split}
\end{equation}
where the poles refer to:
\begin{equation}
\begin{split}
\text{poles}&=\frac{\text{Res} A^s_{\lambda_1\lambda_2\lambda_1\lambda_2}(s=m_{1}^2,0)}{m_{1}^2-s}+\frac{\text{Res} A^s_{\lambda_1\lambda_2\lambda_1\lambda_2}(s=m_{2}^2,0)}{m_{2}^2-s}\\
&+\frac{\text{Res} A^u_{\lambda_1 \bar \lambda_2 \lambda_1 \bar \lambda_2}(u=m_{1}^2,0)}{m_{1}^2-u}
+\frac{\text{Res}A^u_{\lambda_1 \bar \lambda_2 \lambda_1 \bar \lambda_2}(u=m_{2}^2,0)}{m_{2}^2-u}.
\end{split}
\end{equation}

\subsection{Indefinite Scattering}

In order to apply the previous positivity bounds to the two massive spin-2 fields EFTs, we first  calculate the tree-level scattering amplitudes in the forward limit\footnote{The amplitudes were also computed beyond the forward limit in the transversity language but the most stringent bounds are dominated by the forward limit. } ($t=0$) for the interactions given in \eqref{model} for the theory with cycle interactions and in \eqref{action_line} with \eqref{eq:line2diag} and \eqref{cubic_kinetic}-\eqref{line_quartic} for line interactions. We consider three different classes of scattering amplitudes, namely $hh \to hh$, $hf \to hf$ and $ff \to ff$ but the methodology followed for each one is identical:
We decompose the polarization states of the ingoing and outgoing particles in the SVT (scalar, vector, tensor) basis throughout this work (we give the explicit expressions of polarization tensors in \ref{app:pol tens}). The SVT polarizations are essentially real combinations of definite helicity polarizations.  We then apply the positivity bounds stated in \eqref{pos}. In order to find the strongest constraints on the EFT couplings $\{\kappa_3^{(i)}, \kappa_4^{(i)}, c_i,\lambda\}$ and $\{\tilde\alpha_n,\tilde\beta_n\}$ we considered the forward elastic $2-2$ scattering of arbitrary superpositions of the helicity eigenstates. The polarization state of the particle 1 is assumed to be the same as that of the particle 3 and similarly for 2 and~4. Therefore, the entire configuration of helicities of the ingoing and outgoing particles is specified by ten (potentially complex) numbers as\footnote{Strictly speaking two overall normalizations and a phase factor out in the positivity bounds and so there are $2 \times 10-3=17$ independent degrees of freedom. Given this it will not be necessary to normalize the polarization vectors.}
\begin{align}
\label{indefpolhel}
\begin{split}
&\epsilon^{(1)}=\alpha_{T1}\epsilon_{T1}+\alpha_{T2}\epsilon_{T2}+\alpha_{V1}\epsilon_{V1}+\alpha_{V2}\epsilon_{V2}+\alpha_S\epsilon_{S}, \\
&\epsilon^{(2)}=\beta_{T1}\epsilon_{T1}+\beta_{T2}\epsilon_{T2}+\beta_{V1}\epsilon_{V1}+\beta_{V2}\epsilon_{V2}+\beta_S\epsilon_{S},\\
&\epsilon^{(3)}=\epsilon^{(1)},\\
&\epsilon^{(4)}=\epsilon^{(2)}.
\end{split}
\end{align}
The full results for the scattering amplitudes for indefinite scatterings with arbitrary values of $\alpha$'s and $\beta$'s for the case of the cycle theory are given in Appendix~\ref{boundsdef}, where for  simplicity we restrict ourselves to real\footnote{An explicit check showed that complex values did not strengthen the positivity bounds. } coefficients $\alpha$'s and $\beta$'s.\\

In the following subsections we discuss the general constraints that we obtain from the positivity bounds on the various parameters appearing in the action \eqref{model} and in \eqref{action_line}.
In principle one should explore all possible choices of $\alpha$'s and $\beta$'s however we will show below that there is a specific way to pick just a few particular polarization configurations that lead to some of the  strongest bounds on the EFT couplings (specified in Appendix~ \ref{boundsdef}). We shall discuss these and other special cases in detail below.\\

Finally, let us remark that analogous computations can be performed also in the transversity basis \cite{deRham:2017zjm}. In distinction from the SVT basis, there the particle spins are projected in the direction transverse to the interaction plane. In the case of forward scattering it is however sufficient to work in the standard helicity or SVT basis.

\subsection{Summary of Results}\label{sec:summary}
One of the most important results that will be illustrated in specific examples throughout sections~\ref{sec:cycle}, \ref{2metricssupp} and \ref{sec:line} is how the very existence of interactions between the two fields leads to much more stringent bounds on all the parameters of the EFT including on the fields self-couplings, as compared to what one would have had were those two fields entirely isolated from one another.

The naive expectation is that an EFT that contains more fields or fields of higher spins will contain more undetermined coupling constants which implies more free parameters to fit phenomenological data. In reality what has been found already in \cite{deRham:2018qqo} is that this naive argument does not account for the parametrically larger and stronger number of positivity bounds that not only constrain the mixing between the fields (or between the various polarizations of each field) but also the fields self-interactions. Even more importantly, the very existence of an additional pole in the low-energy EFT to be subtracted from the amplitude leads by itself to stronger bounds. This illustrates how demanding a particular field content in the low-energy EFT can be very severely constrained, irrespectively of what the precise interactions are. Turned the other way, this shows how demanding a standard high-energy completion  constrains not only the operators of a low-energy EFT but also the field content and more specifically, the mass and spin distribution of fields with spin-2 or higher.

\section{Bounds for Cycle Interactions}\label{sec:cycle}

In this section we use the general formalism presented above to constrain the EFT coefficients for the theory of two interacting spin-2 fields with cycle interactions presented in section~\ref{sec:cycle_intro}. In particular, we shall compute the forward $2$--$2$ scattering amplitudes between the fields $h_{\mu\nu}$, $f_{\mu\nu}$ described by the action \eqref{model} and impose the constraints implied by the positivity bounds on the dimensionless parameters $\{\kappa^{(1)}_3,\kappa^{(1)}_4,\kappa^{(2)}_3,\kappa^{(2)}_4, c_{1},c_2,\lambda\}$. Together with  the mass ratios $x=m_1/m_2$, $\gamma=M_1/M_2$ and the couplings $d_{1},d_{2}$ (insensitive to the tree level positivity bounds) these form a set of 11 independent dimensionless parameters describing the leading operators of a spin-2 EFT with cycle interactions.

\begin{samepage}
\subsection{General Results}\label{sec:general case}
Here we shall present three general results:
\begin{enumerate}[label=(\roman*)]
\item[\ref{sec:positivity}] The mixed cubic interaction parameters $c_i$ have to be {\bf positive},
\item[\ref{sec:lambda}] The coupling of the {\bf quartic interaction $\lambda \L_{hhff}$ vanishes} $\lambda=0$ in the absence of cubic interactions, \ie when $c_i=0$.
\item[\ref{sec:strongConstraints}] The very existence of an interaction between two fields can lead to {\bf strong constraints even on the fields self-interactions.}
\end{enumerate}
The first two results follow from considering the $hf\to hf$ scattering process with a specific choice of \emph{definite} helicity states and are valid for arbitrary choices of the mass ratios $x = m_1/m_2$ and $\gamma = M_1/M_2$. In general, the strongest constraints from the $hf\to hf$ scattering to the parameters $c_i$ and $\lambda$ are obtained from a finite set ($11$, to be specific) of particular  \emph{indefinite} helicity state choices for the ingoing and outgoing particles. The specific values of $\alpha$'s and $\beta$'s for these configurations are presented in Table~\ref{tab:2} in Appendix~\ref{hfhfSVT}. We also note that for the first two results the $\kappa_4^{(i)}$ self-interactions are irrelevant. The third result already alluded in section~\ref{sec:summary}  follows from considering both mixed $hf\to hf$ and single field $hh\to hh$ (or $ff\to ff$) scattering processes.
\end{samepage}

\renewcommand{\thesubsubsection}{(\roman{subsubsection})}

\subsubsection{Positivity of cubic couplings}\label{sec:positivity}
In what follows we show that the positivity bounds require the cubic mixed couplings to be sign-definite, namely $c_1>0$ and $c_2>0$. This can be concluded from the expression of the forward scattering amplitude given in Eq.~\eqref{indef bound} when requiring that one of the incoming states is a superposition of the two tensor eigenstates, \emph{e.g.} setting $\alpha_{T1},\alpha_{T2}\neq 0$, $\alpha_{V1}=\alpha_{V2}=\alpha_S=0$, while leaving the polarization of the other particle arbitrary. The $hf\rightarrow hf$ bounds for this configuration are given in Table \ref{tab:1} from which we see that the cubic couplings must be positive.

\begin{table}[H]\normalsize
    \centering
    \begin{tabular}{|c|c|c|c|c|c|}
    \hline
         $\alpha_{T1}$ $\beta_{T1}$&$\alpha_{T2}$ $\beta_{T2}$ & $\alpha_{V1}$ $\beta_{V1}$ & $\alpha_{V2}$ $\beta_{V2}$ & $\alpha_{S}$ $\beta_{S}$ & $f$\\
         \hhline{|=|=|=|=|=|=|}
         $\alpha_{T1},\beta_{T1}$ &$\alpha_{T2},\beta_{T2}$ & $0,\beta_{V1}$ & $0,\beta_{V2}$ & $0,\beta_{S}$ & \scriptsize$\left(\alpha _{{T1}}^2+\alpha _{{T2}}^2\right) \left(4 \beta _S^2+3 \left(\beta_{{V1}}^2+\beta_{{V2}}^2\right)\right)$\normalsize$\frac{c_{2}\,m^{2} }{3x^{2}\gamma \Lambda_{3}^{6}}>0$ $\phantom{\Bigg]}$\\[0.2cm]
         \hline
         $\alpha_{T1},\beta_{T1}$ &$\alpha_{T2},\beta_{T2}$& $0,0$ & $0,0$ & $0,0$ & $0$ $\phantom{\Big]}$\\
         \hline
         $\alpha_{T1},\beta_{T1}$ &$\alpha_{T2},\beta_{T2}$& $\alpha_{V1},0$ &$\alpha_{V2},0$&$\alpha_{S},0$ & \scriptsize$\left(\beta _{{T1}}^2+\beta _{{T2}}^2\right) \left(4 \alpha _S^2+3 \left(\alpha_{{V1}}^2+\alpha
   _{{V2}}^2\right)\right)$\normalsize$\frac{c_{1}\,m^{2} }{3x^{2}\gamma \Lambda_{3}^{6}}>0$ $\phantom{\Bigg]}$ \\
    \hline
    \end{tabular}
    \caption{Special configurations of polarizations for $hf\rightarrow hf$ scattering. The first and the third row show that the cubic couplings $c_{i}$ must be positive. The second row indicates that pure tensor states do not impose constraints on the leading order interactions, but could be used to put constraints on contributions from sub-leading operators not included in our EFT. }
    \label{tab:1}
\end{table}

When both of the incoming states are tensors the function $f$ from Eq.~\eqref{pos} is exactly zero. This can be understood by considering the $s$ scaling of different Feynman graphs. By decomposing  the polarization tensors into the polarization vectors,
\begin{equation}\label{poltenvec}
   \epsilon^{\lambda}_{\mu\nu}=\sum_{\lambda'\lambda''}C_{\lambda'\lambda''}^{\lambda}\epsilon^{\lambda'}_{\mu}\epsilon^{\lambda''}_{\nu},
\end{equation}
we can express the scattering amplitude in terms of the following products: $\epsilon^{\lambda}_{(1)\mu}\epsilon^{\lambda'\mu}_{(2)}$,  $\epsilon^{\lambda}_{(1)\mu}p_{2}^{\mu}$,  $\epsilon^{\lambda}_{(2)\mu}p_{1}^{\mu}$ and $p^{\mu}_1 p_{2\mu}$.  By considering the explicit expressions of polarizations and four-momenta (which are given in \eqref{svt vec}-\eqref{svt vec1}), in the forward limit, it can be seen that the only products giving $s$ dependence are $p_1\cdot p_{2}$, $\epsilon^{S}_{(1)}\cdot \epsilon^{S}_{(2)}$, $\epsilon^{S}_{(1)}\cdot p_{2}$ and  $\epsilon^{S}_{(2)}\cdot p_{1}$. Then we choose $\epsilon^{(1)}_{\mu\nu}$ to be a tensor mode, which does not have any $s$ dependence and can be written as \eqref{svt T1} or \eqref{svt T2}, so there can be no factors of $\epsilon^{S}_{(1)\mu}$.

In the $s$-channel diagram both of the cubic vertices are of the double-epsilon structure as in Eq.~\eqref{def_int_2}. The massive spin-2 propagator, $D_{\mu\nu\rho\sigma}(p)$, is given by
\begin{equation}
    D_{\alpha\beta\mu\nu}(p)=-\frac{i \left(G_{\alpha \nu} G_{\beta\mu}+G_{\alpha\mu} G_{\beta\nu}-\frac{2}{3} G_{\alpha\beta} G_{\mu \nu}\right)}{2 \left(m^2+p^2\right)}\,,
\end{equation}
where
\begin{equation}
    G_{\mu \nu}=\eta_{\mu \nu}+\frac{p_{\mu} p_{\nu}}{m^2}\,,
\end{equation}
so $s$ and $u$-channel diagrams which have two $\varepsilon\varepsilon$ vertices can scale at most as $s$ because it is impossible to get more than two factors of $p_1\cdot p_{2}$ and $\epsilon^{S}_{(2)}\cdot p_{1}$ by contracting the indices. Therefore, the $s$ and $u$-channel diagrams do not contribute to the second $s$ derivative. \\

The four point contact interaction diagram gives a sum of products of polarization tensors which can be expressed in terms of products of polarization vectors using Eq.~\eqref{poltenvec}. Since there are no factors of momenta, the only product giving $s$ dependence is  $\epsilon^{S}_{(1)}\cdot \epsilon^{S}_{(2)}$. However, choosing the polarization of particle 1 to be pure tensor again removes all the factors of $\epsilon^{S}_{(1)}$ and, therefore, does not allow any $s$ dependence of this contact interaction diagram. Furthermore, we found that the value of this diagram is exactly zero for a tensor polarization. This leaves the $t$-channel as the only contribution to the second derivative with respect to $s$ because one of the vertices in the $t$-channel is the GR cubic vertex which contains two factors of momenta and hence gives an additional factor of $s$. The constraint shown in Table \ref{tab:1} thus forces the cubic coefficient, $c_{2}$, to be positive. Similarly choosing the other incoming particle to be a pure tensor polarization state forces $c_{1}$ to be positive.

If the second particle is also in the pure tensor state then the only $s$ contribution can come from $p_1\cdot p_{2}$. But there are no such factors in the $t$-channel diagram because the only product of momenta that appears here is $p_1\cdot p_{3}$, so in this case even this diagram does not contribute to the second $s$ derivative and that is why this derivative is exactly zero.

\subsubsection{Vanishing of the Quartic Mixing in the Absence of Cubic Ones} \label{sec:lambda}
Now we consider the simplest special case where there is only the quartic interaction between the two fields, $\L_{hhff}$ defined in \eqref{def_int_2}, \ie we set $c_{2}=c_{1}=0$. We show that this case is not allowed since, in fact, $\lambda$ is also forced to vanish from the positivity bounds. To show this we choose $\beta_{V1}=\alpha_{V2}=\alpha_{S}=0$ and, from the general expression for the $hf\rightarrow hf$ scattering amplitude given in \eqref{indef bound}, obtain the following bound:
\begin{equation}
-18x^6\lambda \left(2 \beta _S^2+2 \sqrt{3} \beta _S \beta _{T1}+3 \beta _{V2}^2\right) \alpha _{V1}^2>0\,.
\end{equation}
The factor $2 \beta _S^2+2 \sqrt{3} \beta _S \beta _{T1}+3 \beta _{V2}^2$ can be both positive or negative for different choices of $\beta$'s and thus implies that in this case $\lambda=0$. This result can be obtained in somewhat more intuitive way from the scattering amplitudes for definite \emph{transversity} eigenstates. Indeed, computing  for instance the bounds coming from the positivity of $f_{1010}$ and $f_{2020}$ in transversity basis \cite{deRham:2017zjm}. In this basis, when $c_1=c_2=0$, we get
\begin{align}\label{F1010}
f^{\rm (trans.)}_{1010}=\frac{\lambda}{12\gamma x^{2}\Lambda_{2}^{4}} >0\,,\qquad{\rm while}\qquad
f^{\rm (trans.)}_{2020}=   -\frac{\lambda}{3\gamma  x^{2}\Lambda_{2}^{4}} >0\,,
\end{align}
and one can directly conclude that $\lambda$ must vanish for $c_i=0$. It follows that in order to have  a Lorentz invariant UV completion, the only pure quartic double-epsilon interactions allowed for the cycle theory of two massive spin-2 fields are $d_1$ and $d_2$.

\subsubsection{Mixing provides Stronger Constraints even for Self--Interactions} \label{sec:strongConstraints} \label{sec:hhhh}

\begin{figure}[h]
    \begin{flushleft}
\includegraphics[height=0.29\textheight]{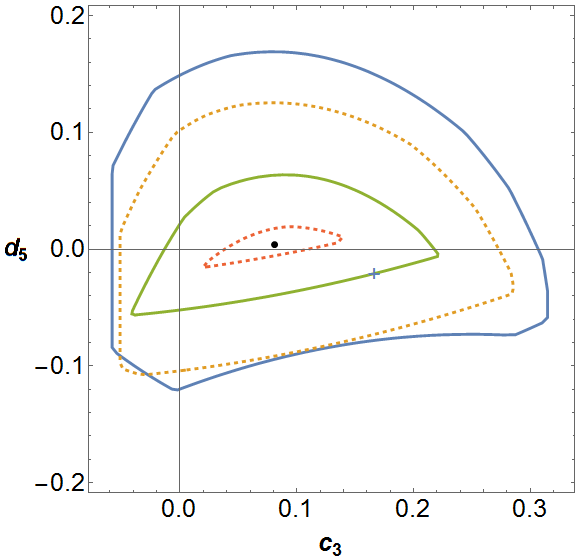} \hspace{0.39cm}
\includegraphics[height=0.29\textheight]{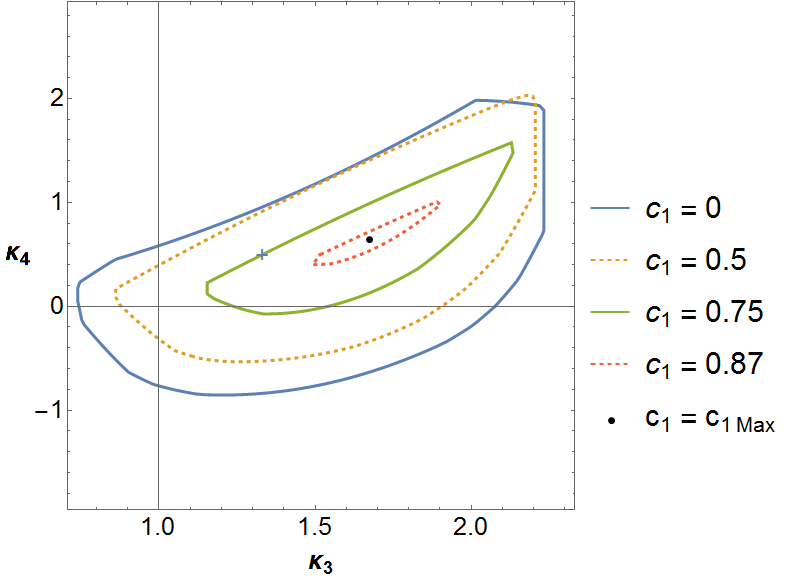}\hspace{-0.5cm}\\
\includegraphics[height=0.29\textheight]{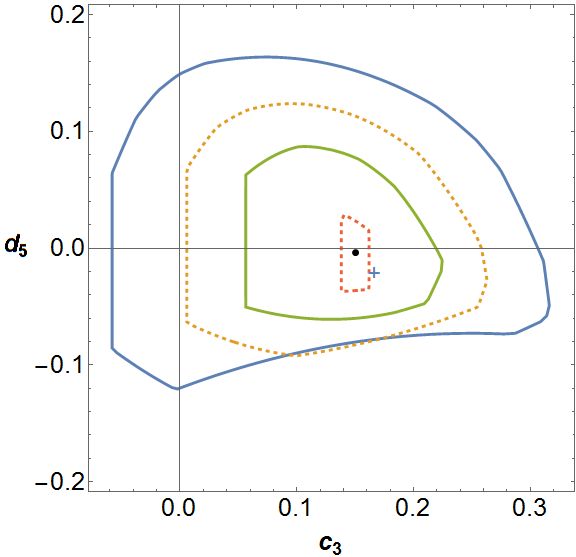}  \hspace{0.39cm}
\includegraphics[height=0.29\textheight]{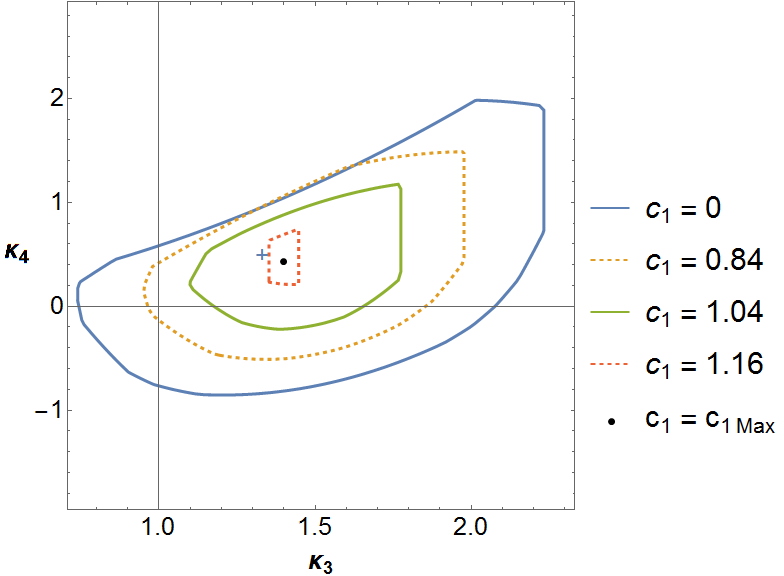}\hspace{-0.5cm}
    \end{flushleft}
\caption{The allowed region of parameters obtained from the indefinite $hh\rightarrow hh$ scattering for different values of $c_1$ at $x=m_1/m_2=0.5$ (up) and $x=2$ (down). The results are presented in both $(c_3,d_5)$ plane (left) and $(\kappa_3,\kappa_4)$ plane (right). By increasing  $c_1$ the island shrinks until it becomes a point at $c_1=c_{1\text{max}}$ shown by the black dot. For $x=0.5$ this point is reached at $c_{1\text{max}}=0.9$, and for $x=2$  at $c_{1\text{max}}=1.2$. The cross in all figures represents the minimal model with $c_3 = 1/6 $ and $d_5 = -1/48$, or $\kappa_3=4/3$, $\kappa_4=1/2$. }
\label{fig:c3d5x.5}
\end{figure}

To illustrate the general result that ``{\it mixing provides stronger constraints even for self--interactions}", we shall investigate the effect of the very existence of mixed interactions  $\mathcal L_{hhf}, \mathcal L_{hff}$ on the allowed parameter region for the self-interactions $\kappa_3^{(i)}$ and $\kappa_4^{(i)}$.
When considering the bounds from the $hh\rightarrow hh$ scattering (and similarly from $ff\rightarrow ff$), the constraints on the self-interactions $\kappa_3^{(i)}$ and $\kappa_4^{(i)}$ have been studied extensively in the earlier literature in the case of a single massive spin-2 field  \cite{Cheung:2016yqr,deRham:2018qqo}. In particular, it is known that there is only a finite  allowed region of parameter space --- an island -- in the self-interaction  parameter space (the two-dimensional $\kappa_3, \kappa_4$  or what is sometimes referred to as the $c_3, d_5$ ghost-free massive gravity parameter space).
In order to enable a direct comparison with the original results of \cite{Cheung:2016yqr}, we cast our results also in the variables $c_3$ and $d_5$ used there following from \cite{deRham:2010ik}. These are related to our parameters used in  \eqref{action} through
\be
\kappa_3^{(i)}=2-4c_3^{(i)}\,,\qquad \kappa_4^{(i)}=1-4c_3^{(i)} - 8d_5^{(i)}\,,\qquad\text{for }\quad i = 1,2\,.
\ee
We drop the superscripts $i$ in the remainder of this section because the overall parameterization is symmetric under $h\leftrightarrow f$ and $1\leftrightarrow 2$. Hence, although we focus our attention on the $hh\to hh$ scattering here, the results for the $ff\to ff$ scattering will give the same bounds on its corresponding mass parameters.

\begin{figure}[H]
    \centering
\includegraphics[width=0.8\textwidth]{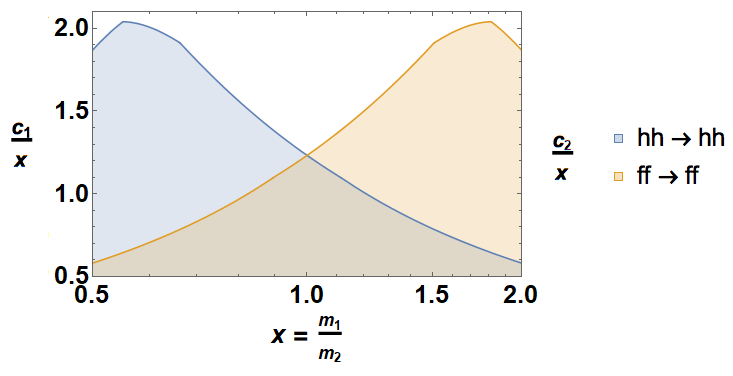}
\caption{The allowed values of the cubic couplings, $c_1$ (blue) and $c_2$ (yellow), as a function of the mass ratio, $x$, obtained from $hh\rightarrow hh$ scattering. For a given value of $x$, the maximal allowed value, $c_1=c_{1\text{max}}$, is determined as the value at which the allowed $(c_3,d_5)$ island shrinks to a point.}
\label{fig:c1vsxhh}
\end{figure}

\begin{figure}[h]
    \centering
\includegraphics[width=0.6\textwidth]{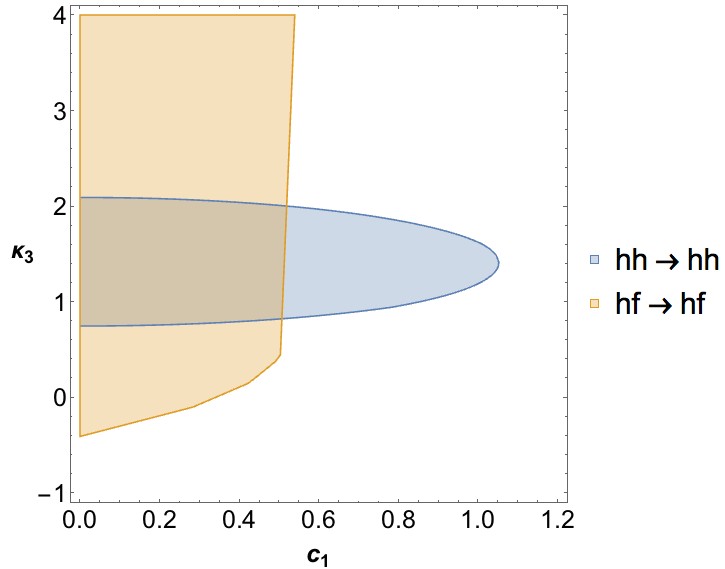}
\caption{Comparison of the allowed region of cubic parameters $c_1, \kappa_3$ obtained from the indefinite $hh\rightarrow hh$ and $hf\rightarrow hf$ scatterings, for $x=m_1/m_2=1$, $c_2=c_1$, and vanishing quartic couplings $\lambda=\kappa_4=0$, $\gamma=1$. Each channel allows the removal of a different region of parameter space.}
\label{fig:c3vsc1}
\end{figure}

        Now if we include a mixing between the two fields $h$ and $f$, this coupling affects the $hh\to hh$ (and $ff\to ff$) amplitudes in a way which is severely constrained by the positivity bounds.
As before, the strongest bounds on the couplings can be extracted from a finite set of specific indefinite helicity polarization choices for the ingoing and outgoing states. These are presented in Table~\ref{tab:3} in Appendix~\ref{hhhhSVT}. The positivity bounds on the $hh\to hh$ scattering amplitudes receive corrections from the exchange diagrams of the $f$ field through the $c_1\mathcal L_{hhf}$ interaction and give a contribution to the scattering amplitude that is proportional to $c_1^2$. The modified allowed parameter region is plotted in $(c_3,d_5)$ and in $(\kappa_3,\kappa_4)$ planes for several different values of $c_1 \ge 0$ (given the results we derived in point~\ref{sec:positivity}) in Figs.~\ref{fig:c3d5x.5}  for the mass ratios $x = m_1/m_2=0.5$ (up) and $x = 2$ (down). We set $\gamma = 1$ for simplicity.\footnote{\label{foot:gamma}We note however that the $hh\to hh$ scattering amplitude is independent on the actual value of $\gamma$. This is apparent from the exact expression of the scattering amplitude given in \eqref{hhhhSVT}.} We find that for a fixed value of $x$ the parameter island \emph{shrinks} as we increase the value of $c_1$. For some maximal value of $c_1 = c_{1\text{max}}(x)$ the allowed parameter region shrinks to a point. This allows us to put an upper bound on $c_1$; we plot this in Fig.~\ref{fig:c1vsxhh}.
For $c_1 =0$ we recover the Cheung-Remmen \cite{Cheung:2016yqr} parameter island (see blue curve in Fig.~\ref{fig:c3d5x.5}). We also find that the value $c_1 = 0$ always gives the maximal range of the allowed values of $c_3$ and $d_5$ (independent on $x,\gamma$) and is given by
\begin{equation}\label{rangec3d5}
-0.0582< c_3<0.315\,,\qquad-0.121< d_5<0.169 \,.
\end{equation}
This is in agreement with the earlier findings of \cite{deRham:2018qqo}. For future reference, we also give the allowed range of $\kappa_3$ and $\kappa_4$:
\begin{equation}\label{rangek3k4}
0.74<\kappa_3<2.23\,,\qquad-0.87<\kappa_4<1.99 \,.
\end{equation}
Finally, let us remark that the so-called minimal model with $c_3=1/6$ and $d_5 = -1/48$, or $\kappa_3=4/3$, $\kappa_4=1/2$ corresponds to a theory with no decoupling limit interactions in the helicity-0/helicity-2 sector in the one-field case \cite{deRham:2010ik}. This specific case can always be ruled out for sufficiently high values of $c_1$, but we also emphasize that the role of that model is of somewhat limited value when including mixings between the two spin-2 fields since other interactions are then present in the decoupling limit. Nevertheless, to connect with the previous literature
 we denote the minimal model with a cross in Figs.~\ref{fig:c3d5x.5} and all other figures where relevant below. As we shall see in the next subsection, when including the positivity bounds from the $hf\to hf$ scattering, there is also a bound on how large $c_1$ can be and for values of $c_1$ within that bound, the minimal model remains in the allowed region of parameter space. The constrains from the $hh\to hh$ and $hf \to hf$ channels work in a very  complementary way and allow to remove different regions of parameter space as is illustrated in Fig.~\ref{fig:c3vsc1} where the constraints from $hf\to hf$ are compared to those arising from $hh\to hh$ in the $\{c_1,\kappa_3\}$ plane, in the $\mathbb{Z}_2$-symmetric case at $\lambda=\kappa_4=0$.

\subsection{$\mathbb{Z}_2$ Symmetric Case}\label{sec:z2}

\begin{figure}[H]
    \centering
\includegraphics[width=0.45\textwidth]{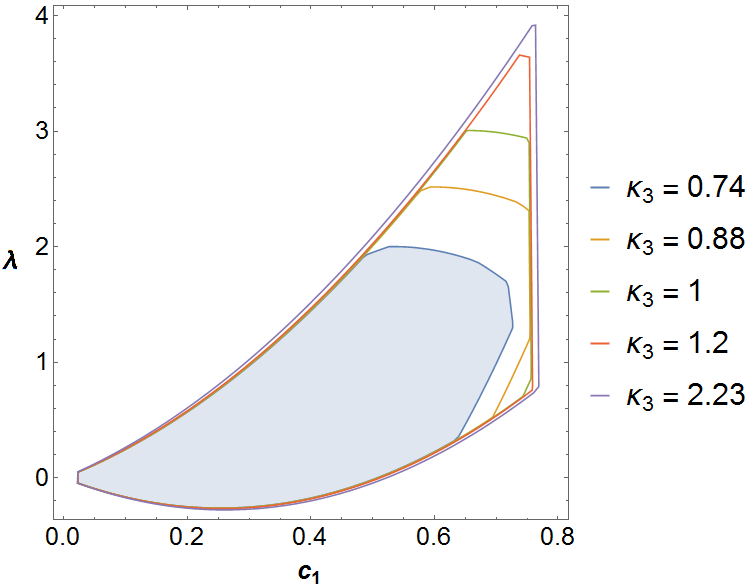}\qquad
\includegraphics[width=0.475\textwidth]{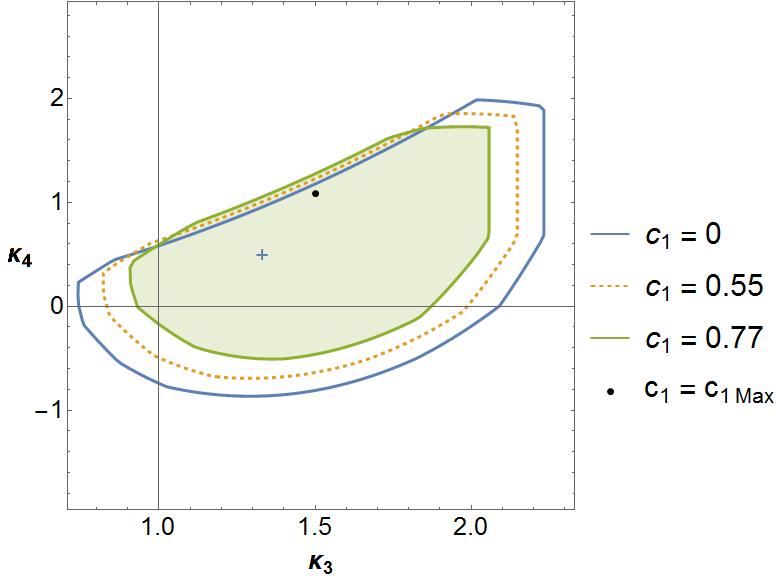}
\caption{Relation between the bounds from $hh\to hh$ and $hf\to hf$. \\
{\bf Left:} The allowed region of parameters $(\lambda,c_1)$ obtained from combining positivity bounds from both $hh\rightarrow hh$ and $hf\rightarrow hf$ scatterings in the $\mathbb{Z}_2$ symmetry case. The largest allowed parameter region is obtained from $\kappa_3=2.23$ and the smallest --- from $\kappa_3=0.74$ (shaded). In all cases we recover $\lambda=0$ when $c_1=0$. \\
{\bf Right:} The allowed region of $(\kappa_3,\kappa_4)$ obtained from the $hh\rightarrow hh$ scattering in the $\mathbb Z_2$~symmetric case. The island starts to shrink as $c_1$ increases until it reaches  $c_{1\text{max}}^{\mathbb Z_2}=0.77$ (green). The bounds obtained from the $hf\rightarrow hf$ scattering amplitudes forbids the shrinking of the island to a point which would occur at $c_{1\text{max}}=1.23$. The minimal model with $\kappa_3=4/3$, $\kappa_4=1/2$ is depicted by a cross.}
\label{fig:c3d5x.1}
\end{figure}

In general we are dealing with a nine--dimensional parameter space $x, \gamma, \kappa_{3,4}^{(1,2)}, c_{1,2}, \lambda$ and providing the generic positivity constraints in the full nine--dimensional space is beyond the scope of this work. However much progress can be made by investigating specific slices of this nine--dimensional manifold.  One of the most natural scenarios to consider is the one that enjoys a  $\mathbb{Z}_2$ symmetry with respect to swapping the two fields, $h$ and $f$, corresponding to $c_{1}=c_{2}$, $\kappa^{(1)}_{n}=\kappa^{(2)}_{n}$, $x=m_1/m_2=1$ and $\gamma=M_1/M_2=1$. From the previous results obtained from positivity bounds on the indefinite $hh\rightarrow hh$ scattering, shown in Fig.~\ref{fig:c1vsxhh}, we find the allowed range of the cubic coupling $c_1$ (for $x=1$):
\begin{equation}
    0< c_1<1.23\,.
\end{equation}
By combining these positivity bounds from the indefinite $hh\rightarrow hh$ scattering with the bounds from the indefinite $hf\rightarrow hf$ scattering, we get further constraints on the parameters $\lambda$ and $c_1$:
\begin{equation}
\begin{split}
    0<&\,\,c_1<0.77\,,\\
    -0.28<&\,\,\lambda<3.93\,.\\
\end{split}
\end{equation}
The full allowed parameter space in the $(\lambda,c_1)$ plane is plotted in Fig.~\ref{fig:c3d5x.1} [Left] for different values of $\kappa_3$. We see that the allowed region \emph{shrinks} as $\kappa_3$ decreases from its maximum, $\kappa_3=2.23$, to its minimum at $\kappa_3=0.74$. We also note that including the bounds from the $hf\to hf$ scattering shifts the maximal allowed value of the cubic interaction, $c_1$, from $c_{1\text{max}}=1.23$ to $c_{1\text{max}}^{\mathbb Z_2}=0.77$. As discussed in the previous section, by increasing $c_1$ until  $c_{1\text{max}}=1.23$, the allowed $(\kappa_3,\kappa_4)$ island (or, equivalently, the $(c_3, d_5)$ island) from the $hh\to hh$ scattering shrinks to a single point. By including the  new bounds from $hf\rightarrow hf$ one is not allowed to increase $c_1$ beyond $c_{1\text{max}}^{\mathbb Z_2}=0.77$ thus preventing the island from disappearing. The $(c_3,d_5)$ island for different values of $c_1$ is shown in Fig.~\ref{fig:c3d5x.1} [Right]. We remark that also the minimal model with $c_3=1/6$, $d_5=-1/48$, or $\kappa_3=4/3$, $\kappa_4=1/2$ (the green cross in Fig.~\ref{fig:c3d5x.1} [Right]) is still allowed by the combined positivity bounds.

\subsection{\texorpdfstring{$\mathbb{Z}_2$ Symmetric Case with  $x\neq1$}{$\mathbb{Z}_2$ symmetric case with, xneq1}}
\begin{figure}[H]
    \centering
\includegraphics[width=0.7\textwidth]{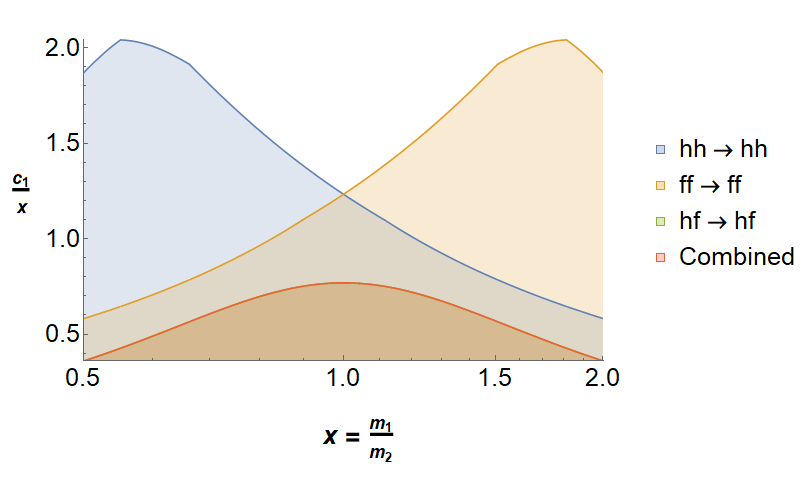}
\caption{The allowed values of the cubic coupling, $c_1$, as a function of the mass ratio, $x$, for the case where all the couplings are equal and $\gamma=1$. The allowed region is obtained from $hh\rightarrow hh$, $ff\rightarrow ff$  and  $hf\rightarrow hf$ scattering. The maximal value for $c_1/x$ in the $\mathbb{Z}_2$ case is at $x=1$ and corresponds to 0.77.  On the other hand, the maximum value of the cubic coupling $c_1$ occurs at $x=1.32$ and is $c_{1\text{max}}=0.86$.
 Positivity bounds from the mixing channels $hf\to hf$ are more constraining than those from the single-field channels $hh\to hh$ and $ff\to ff$. }
\label{fig:c1vsxz2x}
\end{figure}

Now we consider a slightly more general case with $\kappa_n^{(1)}=\kappa_n^{(2)}$, $c_1=c_2$ and $\gamma=1$ while letting $x\neq1$. In this case we get the maximum value of $c_1$ as a function of $x$ from the combined positivity bounds on the $hh\rightarrow hh$, $ff\rightarrow ff$  and  $hf\rightarrow hf$ scattering amplitudes. This is shown in Fig.~\ref{fig:c1vsxz2x}. We note that since we demand that $c_1=c_2$ the bounds from $hh\to hh$ and $ff\to ff$ now need to be combined in this case (compared to Fig.~\ref{fig:c1vsxhh}). Also, we see that there are new bounds on $c_1$ coming from the $hf\rightarrow hf$ process. These appear because the allowed $(\lambda,\kappa_3^{(1)})$ region \emph{shrinks} for increasing values of~$c_1$ leading to a maximal allowed value of $c_1$ such that allowed region is not empty. The absolute maximum value (from those combined bounds) for the cubic coupling in this case is $c_{1\text{max}}=0.86$ occurring at $x=1.32$.
The couplings $c_1$ and $\lambda$ are bounded as follows:
\begin{equation}\label{quarticcubibc}
\begin{split}
    -0.37<& \,\,\lambda<4.64\, ,\\
    0<& \,\,c_1 <0.86\,.
\end{split}
\end{equation}

\subsection{\texorpdfstring{$\mathbb{Z}_2$ Symmetric Case with  $\gamma\neq1$}{$\mathbb{Z}_2$ symmetric case with, xneq1}}
\begin{figure}[H]
    \centering
\includegraphics[width=0.6\textwidth]{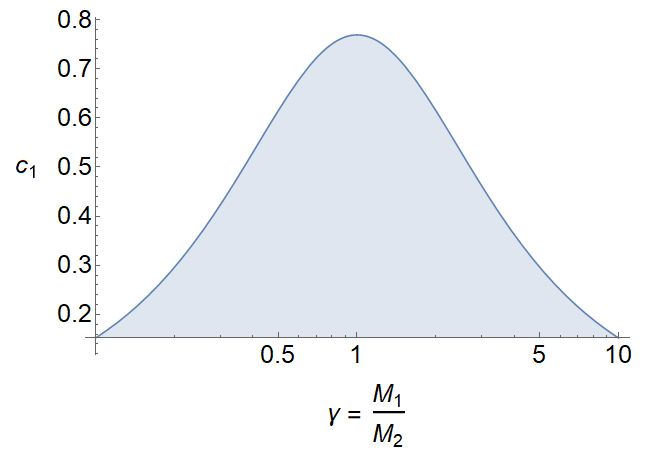}
\caption{The allowed values of the cubic coupling, $c_1$, obtained from the $hf\rightarrow hf$ scattering as a function of $\gamma$ for the case where all the other couplings are taken to be $\mathbb Z_2$ symmetric (and in particular $x=1$). The maximum value for the cubic coupling occurs at $\gamma=1$ and is again $c_{1\text{max}}^{\mathbb Z_2}=0.77$. }
\label{fig:c1vsxz2gamma}
\end{figure}

Let us now consider a similar case with $\kappa_n^{(1)}=\kappa_n^{(2)}$, $c_1=c_2$ and $x=1$ while letting $\gamma\neq1$. In this case the maximum value of $c_1$ as a function of $\gamma$ is determined solely from the positivity bounds on the $hf\rightarrow hf$ scattering. The reason for this is that, as already mentioned in footnote \ref{foot:gamma}, the bounds on the $hh\to hh$ (and similarly on the $ff\to ff$) scattering are independent on the value of $\gamma$. In particular, for $x = 1$ we found in the $\mathbb Z_2$ symmetric case that the maximal value of $c_1$ from the $hh \to hh$ scattering is $c_{1\text{max}}=1.23$. This is a much weaker bound than the one coming from the $hf\to hf$ scattering. The latter results are shown in Fig.~\ref{fig:c1vsxz2gamma}. We find that the absolute maximum value for the cubic coupling, $c_1$, in this case occurs at $\gamma = 1$ and equals to $c_{1\text{max}}^{\mathbb Z_2}=0.77$. As expected, this coincides with the maximal value of $c_1$ found in the $\mathbb Z_2$ symmetric case in section~\ref{sec:z2}. The couplings $c_1$ and $\lambda$ are bounded as follows:

\begin{equation}\label{gamma}
\begin{split}
    -0.28<& \,\,\lambda<3.93\, ,\\
    0<& \,\,c_1 <0.77\,,
\end{split}
\end{equation}
which is comparable (in spirit) to the bounds found in \eqref{quarticcubibc}.

\subsection{One Cubic Interaction}
As a last concrete example, let us consider the positivity bounds from the $hf\to hf$ scattering with a single cubic interaction, $c_1 \L_{hhf}$,  and the quartic interaction, $\lambda \mathcal L_{hhff}$. In other words, we set the coefficient of $c_2 \L_{hff}$ to  $c_2=0$. This will allow us to constrain the cubic and quartic couplings $c_1$ and $\lambda$. We set $\gamma=1$ for simplicity and note that the mass terms $\kappa_4^{(i)}$ and $\kappa_3^{(1)}$ do not contribute to this process. For this case the constraints on $c_1$ from the $hh\to hh$ channel (to which $\kappa_4^{(i)}$ and $\kappa_3^{(1)}$ do contribute) have already been computed and are identical to the case when $c_2\ne 0$. As for the channel $ff\to ff$, this provides no constraint on $c_1$ (when $c_2\ne c_1$). \\

\begin{figure}[H]
    \centering
\includegraphics[width=0.7\textwidth]{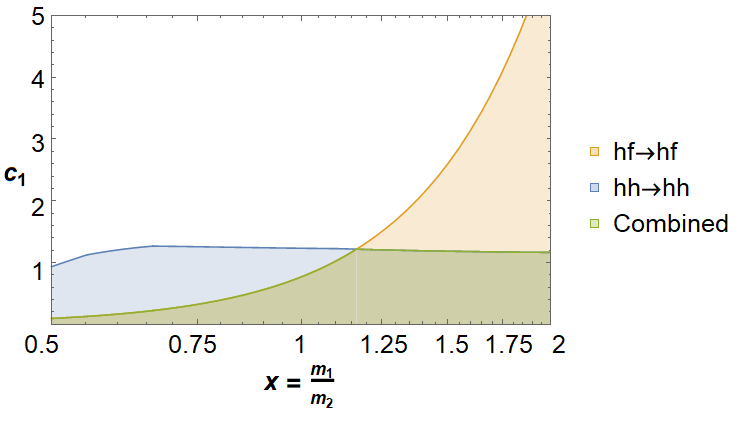}
\caption{The allowed values of the cubic coupling, $c_1$, as a function of the mass ratio, $x$, for $c_2=0$ and $\gamma=1$. The maximum values of $c_1$ (solid lines) correspond to the $c_1$ values at which the allowed island in the $(\kappa_3,\kappa_4)$ plane shrinks to a point (blue) or the bounds from $hf\rightarrow hf$ scattering cannot be satisfied (orange). The combined maximum value of $c_1$ is $c_{1\text{max}}^{\text{one cubic}}=1.22$ occurring at $x=1.17$.}
\label{fig:xVSc1singlecubic}
\end{figure}

We find that the maximal allowed region for the couplings is obtained by setting $\kappa_{3}^{(2)}$ to its maximum value, $\kappa_{3}^{(2)}=2.23$, from Eq.~\eqref{rangek3k4}.  By combining the bounds from the  $hf\rightarrow hf$ scattering with the ones from $hh\rightarrow hh$ scattering, the cubic and quartic couplings are bounded as follows:
\begin{equation}\label{quartic min}
\begin{split}
    -0.72<& \,\,\lambda<3.27\, ,\\
    0<& \,\,c_1 <1.22\,.
\end{split}
\end{equation}
The allowed region of $c_1$ is plotted in Fig.~\ref{fig:xVSc1singlecubic} as a function of $x$. We emphasize that in this case, the maximal allowed value of $c_1$ is given by $c_{1\text{max}}^{\text{one cubic}}=1.22$ that is larger than the maximal value allowed in the $\mathbb Z_2$ symmetric case, $c_{1\text{max}}^{\mathbb Z_2}=0.77$. Hence, in the case with only one cubic interaction one can access larger values of $c_1$ thus enabling a further shrinking of the allowed $(\kappa_3,\kappa_4)$ parameter region as described in subsection~\ref{sec:hhhh}.

\section{$\Lambda_3$ Cycle Theory}\label{2metricssupp}
In \cite{Alberte:2019lnd} it was found that the strong coupling scale of the theory with cycle interactions \eqref{model} can be raised from $\Lambda_{7/2}=(m^{5/2}M)^{2/7}$ to $\Lambda_3 = (m^2 M)^{1/3}$ by performing the following rescaling of the interaction parameters
\be\label{scaling}
c_1\to c_1\frac{m}{\Lambda_{3}}\,,\quad c_2\to c_2\frac{m}{\Lambda_{3}}\,,\quad\lambda\to\lambda\frac{m}{\Lambda_3}\, , \quad  d_1\to d_1\frac{m}{\Lambda_{3}}\,,\quad d_2\to d_2\frac{m}{\Lambda_{3}} \,.
\ee
When compared to the coupling constants of the self-interactions in \eqref{model} this implies in particular that $c_{1,2},\lambda, d_{1,2} \ll \kappa_n^{(i)}\sim \mathcal O(1)$. The resulting EFT is the ultimately highest possible cutoff EFT for two massive spin-2 particles with cyclic interactions.

The aim of this section is to impose the scaling \eqref{scaling} on our parameters only partially and check whether positivity bounds show any preference towards the parameter space region leading to the higher strong coupling scale. In practice, since the ratio $m/\Lambda_3\ll1$, the full scaling \eqref{scaling} amounts to setting $c_i,\lambda\ll1$ in the positivity bounds while keeping the mass parameters $\kappa_n^{(i)}\sim\mathcal O(1)$. Let us now consider the possibility that only the cubic interactions $c_i$ between the two spin-2 fields are suppressed while keeping the coupling of the quartic interactions, $\lambda$, unconstrained. In what follows we thus consider $c_{1,2}\ll \kappa^{(i)}_{n}$, and ignore the quadratic terms $c_{1,2}^2\ll c_{1,2}$. We first note that, according to \eqref{Island}, the corrections to the $hh\rightarrow hh$ scattering amplitudes in our theory relative to the theory of a single massive spin-2 field, $h$ or $f$, are proportional to $c_1^2$ or $c_2^2$ which we effectively set to zero in this case. Hence, the original ghost-free massive gravity constraints on $\kappa_3$'s, obtained  in \cite{Cheung:2016yqr,deRham:2018qqo} from this scattering process, still apply:
\begin{equation}\label{islandrange}
\begin{split}
 0.741<&\,\kappa^{(i)}_{3}<2.23\,.
\end{split}
\end{equation}
Due to the suppression of cubic interactions these results are valid for arbitrary values of $\gamma, x$.

\begin{figure}[t]
    \centering
\includegraphics[height=5.5cm]{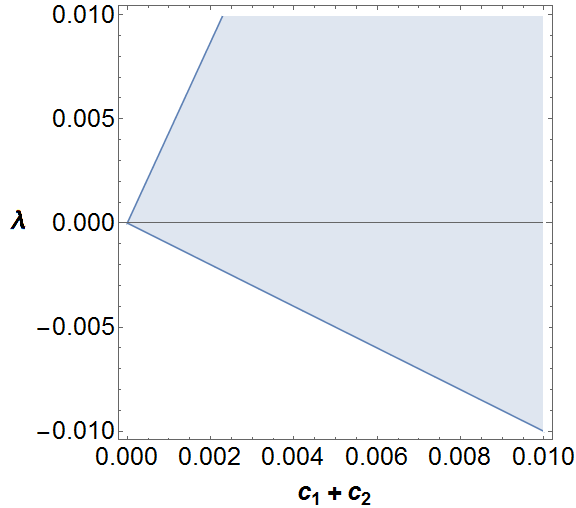}
\qquad
\includegraphics[height=5.5cm]{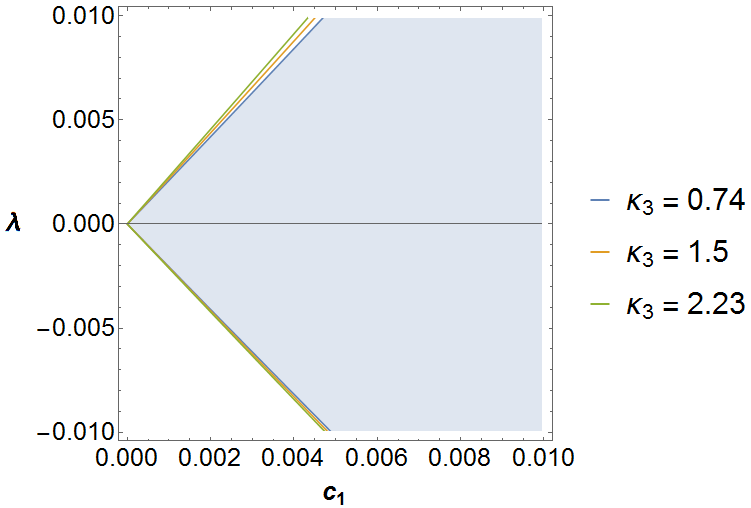}
\caption{The allowed regions in the case of  suppressed cubic interactions. \\
\textbf{Left:} General case for arbitrary values of $x=m_1/m_2$ and $\gamma=M_1/M_2$. \\
\textbf{Right:} $\mathbb{Z}_2$ symmetric case, \ie with $x=\gamma=1$. In both cases the magnitude of the quartic coupling $\lambda$ is bounded by $c_1$ and $c_2$. Therefore it also must go to zero when $c_1,c_2\rightarrow0$ in order to be consistent with the positivity bounds. }
\label{fig:Supp}
\end{figure}

By considering special configurations of the polarizations of the ingoing and outgoing particles one can put upper and lower bounds on the values of $\lambda$. In particular, a configuration where all $\alpha$'s and $\beta$'s are zero, except for $\alpha_{V1}$ and $\beta_{V2}$ gives an upper bound on $\lambda$ from Eq.~\eqref{indef bound} as:
\begin{equation}\label{quartic constr}
    \lambda<\bigg(1+\frac{3}{2}\kappa_3^{(1)}\bigg)c_2+\bigg(1+\frac{3}{2}\kappa_3^{(2)}\bigg)c_1\,.
\end{equation}
Using the fact that $c_{1,2}>0$ and substituting the maximum values for $\kappa_3$'s from \eqref{islandrange} gives:
\begin{equation}\label{lambdamax}
    \lambda<4.35(c_1+c_2)\,.
\end{equation}
Now choosing the non-zero $\alpha$'s and $\beta$'s to be $\alpha_{T1}=\beta_{T1}$ and $\alpha_{S}=\beta_{S}\ll\alpha_{T1}$ we get
\begin{equation}\label{lambdamin}
    \lambda>-(c_1+c_2)\,.
\end{equation}
Combining the two bounds leads to the allowed region for $\lambda$ (shown in Fig.~\ref{fig:Supp}):
\begin{equation}
    -(c_1+c_2)<\lambda<4.35(c_1+c_2)\,.
\end{equation}
We note that these are not be the strongest constraints on the allowed values of $\lambda$. However, they provide robust analytic bounds and necessary constraints implying that the order of magnitude of $\lambda$ has to be close to that of the cubic couplings, $c_i$. A similar but more constraining bound can be obtained when the cubic couplings $c_{1,2}$ are suppressed in the case with $\mathbb{Z}_2$ symmetry. The allowed region in $(c_1,\lambda)$ plane for different values of $\kappa_3$ is also shown in Fig.~\ref{fig:Supp} (right). We can see that the magnitude of $\lambda$ is bounded by a contribution linear in $c_1$.

The previous result is reminiscent to what happens in the theory of a single massive spin-2 field. There positivity bounds, in particular non-forward limit ones \cite{deRham:2017zjm}, impose the tunings, at least to quartic order, that raise the cutoff from $\Lambda_5$ to $\Lambda_3$. In other words tree level positivity bounds enforce the dRGT mass terms with the double-epsilon structure to the order up to which they contribute to the scattering amplitudes \cite{deRham:2018qqo}.

\section{Bounds for Line Interactions}\label{sec:line}
In this section we apply the positivity bounds on the $2-2$ scattering amplitudes in the theory of two interacting massive spin-2 fields with line interactions. The theory is described by the action \eqref{eq:actionline} and the perturbative Lagrangian~\eqref{action_line} with its various components given in Eqns. (\ref{eq:line2diag}--\ref{line_quartic}). Note that although the various mass and non-derivative interactions are almost identical in structure to the case of the cycle theory (see Table~\ref{tab:compare}) the main distinction lies in the non-trivial mixing occurring in the cubic and quartic kinetic terms \eqref{cubic_kinetic}, \eqref{quartic_kinetic}. Also, in comparison to the cycle theory, there are only~6 (as opposed to 11) independent dimensionless parameters $\{\tilde\alpha_3,\tilde\alpha_4,\tilde\beta_3,\tilde\beta_4, \tilde x=\tilde m_1/\tilde m_2,\gamma=M_1/M_2\}$ for the leading-order operators in this EFT.

\begin{figure}[b]
    \centering
\includegraphics[height=8cm]{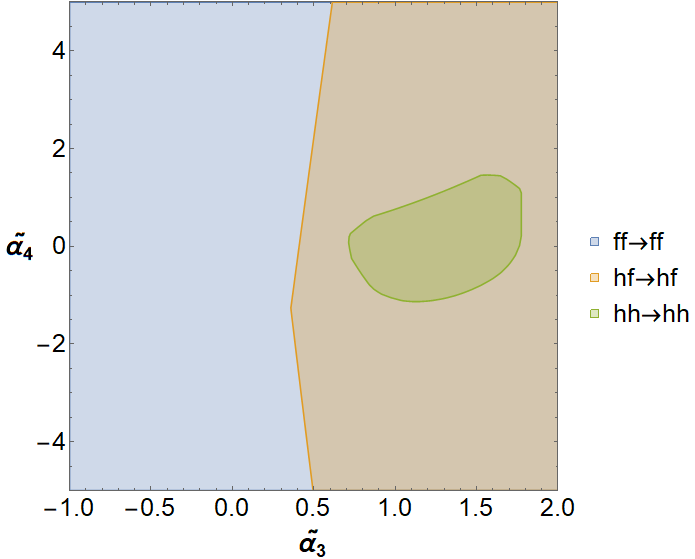}
\caption{The allowed regions in the $(\tilde \alpha_3,\tilde \alpha_4)$ plane in the case of $\tilde \beta_3=-1, \tilde \beta_4=0$ for the mass ratios $\tilde x=2.2,\gamma=1$ (corresponding to a mixing angle $\theta=0.1$). For these (generic) allowed values of $(\tilde \beta_3,\tilde \beta_4)$ the constraints on the couplings $(\tilde \alpha_3,\tilde \alpha_4)$ come mainly from the $hh\rightarrow hh$ scattering. }
\label{fig:a3a4}
\end{figure}

\subsection{General Results}\label{sec:line_general}
\begin{figure}[h]
    \centering
\includegraphics[height=8cm]{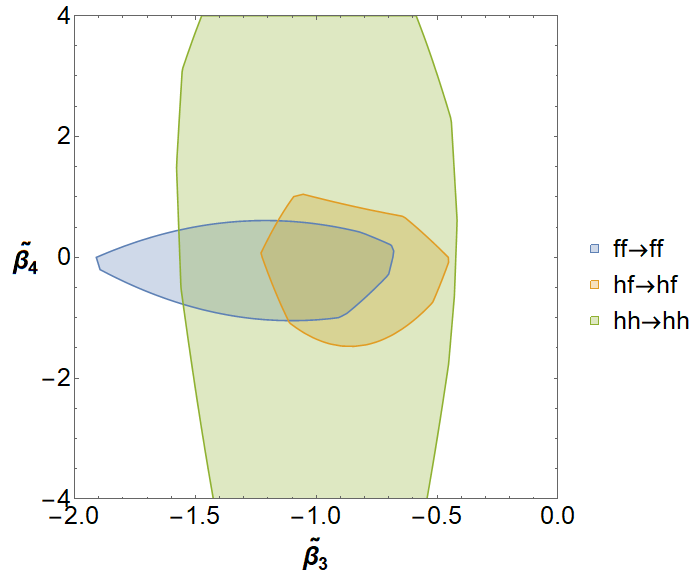}
\caption{The allowed regions in the $(\tilde \beta_3,\tilde \beta_4)$ plane in the case of $\tilde \alpha_3=1, \tilde \alpha_4=0$ for the mass ratios $\tilde x=2.2,\gamma=1$ (corresponding to a mixing angle $\theta=0.1$). For the (generic) allowed values of $(\tilde \alpha_3,\tilde \alpha_4)$ the constraints on $(\tilde \beta_3,\tilde \beta_4)$ come mainly from the $ff\rightarrow ff$ and $hf\rightarrow hf$ scatterings.}
\label{fig:b3b4}
\end{figure}

The generic parameter regions allowed by the combined positivity bounds imposed on the $hh\rightarrow hh$, $hf\rightarrow hf$ and $ff\rightarrow ff$ forward scattering processes are shown in Fig. \ref{fig:a3a4} and Fig. \ref{fig:b3b4}. We see that for some characteristic allowed values of $(\tilde \beta_3,\tilde \beta_4)$ the strongest constraints on the couplings $(\tilde \alpha_3,\tilde \alpha_4)$ come mainly from the $hh\rightarrow hh$ scattering. In turn, the strongest constraints on $(\tilde \beta_3,\tilde \beta_4)$ come mainly from the $ff\rightarrow ff$ and $hf\rightarrow hf$ scatterings. It is important to note that this result is obtained for a mixing angle between the interaction eigenstates, $\theta = 0.1$, in which case $\tilde h_{\mu\nu}\approx h_{\mu\nu}$ and $\tilde f_{\mu\nu}\approx f_{\mu\nu}$. The result is then not surprising, given the line action \eqref{eq:actionline} where the couplings $(\tilde \alpha_3,\tilde \alpha_4)$ control the self-interactions of $\tilde h_{\mu\nu}$ while $(\tilde \beta_3,\tilde \beta_4)$ are responsible for the self-interactions of $\tilde f_{\mu\nu}$ and the mixing between $\tilde h$ and $\tilde f$.

\begin{figure}[h]
    \centering
\includegraphics[height=8cm]{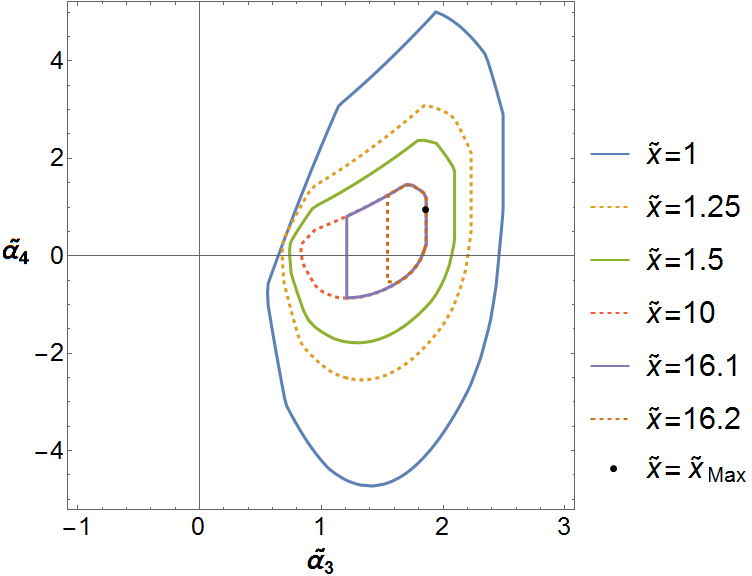}
\caption{The allowed regions in $(\tilde \alpha_3,\tilde \alpha_4)$ plane in the case of $\gamma=1$ and $\tilde \beta_3=-0.77, \tilde \beta_4=-0.06$. The island shrinks to a point at $(\tilde \alpha_3=1.85,\tilde \alpha_4=0.95)$ for the maximum value of $\tilde x_{\rm max}=16.29$. For these values of $\gamma, \tilde x$ the constraints on $(\tilde \alpha_3,\tilde \alpha_4)$ come mainly from the $hh\rightarrow hh$ scattering.}
\label{fig:13}
\end{figure}

In Fig.~\ref{fig:13} we show the allowed parameter region of $(\tilde \alpha_3,\tilde \alpha_4)$ for $\gamma =1$ and varying $\tilde x$. The values of $(\tilde \beta_3,\tilde \beta_4)$ are chosen in order to maximize the allowed parameter space. The allowed region shrinks as we increase $\tilde{x}\geq 1$ and becomes a point at $\tilde{x}_{\rm max}=16.29$ (corresponding to the physical mass ratio $x=m_1/m_2=23.08$). This is equivalent to changing the mixing angle in the range from $\theta = [\pi/8,0.002]$ thus explaining why the main constraints still arise from the $hh\to hh$ scattering as above.

\begin{figure}[h]
    \centering
\includegraphics[height=7cm]{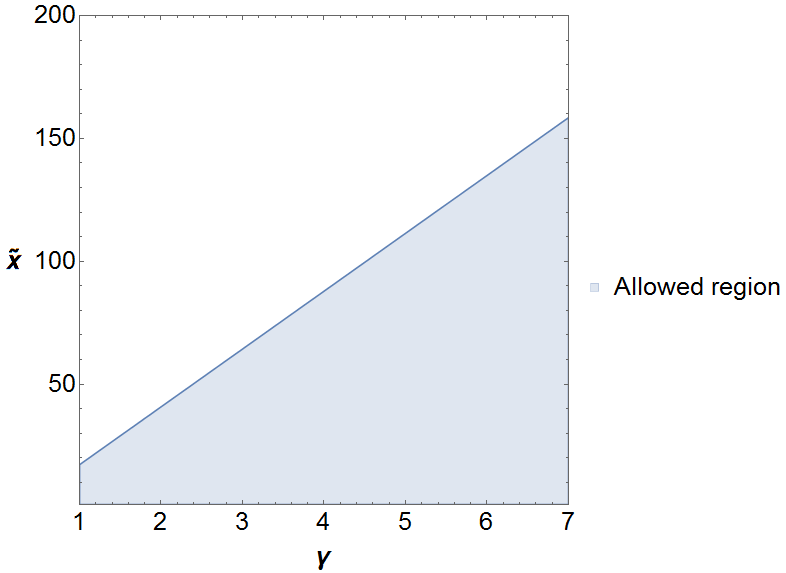}
\caption{The allowed parameter region in $(\tilde x,\gamma)$ plane obtained from the combined constraints from all three scattering processes. For any fixed value of $\gamma$, the maximal value for $\tilde x$ is obtained  by increasing $\tilde x$ until the allowed region in the parameter space of $\tilde \alpha_3$, $\tilde \alpha_4$, $\tilde \beta_3$ and $\tilde \beta_4$  shrinks to a point. The blue line represents the maximum value of $\tilde x$ for a given $\gamma$. A linear relation between $\tilde x_{\rm max}$ and $\gamma$ is to be expected and is directly related to the presence of another spin-2 pole.}
\label{fig:xt vs gamma}
\end{figure}

We also find the maximum value $\tilde x_{\rm max}$ beyond which the allowed parameter region in the full $(\tilde \alpha_3,\tilde \alpha_4,\tilde \beta_3,\tilde\beta_4)$ parameter space shrinks to zero for several values of $\gamma\neq 1$. We do so by combining the results from all three scattering processes. We then numerically manipulate the three-dimensional parameter subspace $(\tilde \beta_3,\tilde \beta_4,\tilde \alpha_3)$ for various values of $\tilde\alpha_4$ and determine $\tilde x_{\rm max}$ as the value beyond which the allowed three-dimensional $(\tilde \beta_3,\tilde \beta_4,\tilde \alpha_3)$ region is void for all values of $\tilde\alpha_4$. The result is shown in Fig.~\ref{fig:xt vs gamma}.  Interestingly the maximal allowed value for $\tilde x_{\rm max}$ is linearly proportional to $\gamma$, which is directly related to the existence of the additional pole of the other spin-2 field as will be clarified below.

\subsection{Decoupling}
In this subsection we establish the decoupling limit in which we recover the theory of a single massive spin-2 field from the theory \eqref{eq:actionline} of two interacting massive spin-2 fields with line interactions. We then investigate the departures from the known results for a single massive spin-2 field \cite{Cheung:2016yqr}.

We consider the limit when
\be
\tilde m_1\to \infty\,,\quad M_1\to \infty\,,\qquad\text {and}\quad \tilde m_2\equiv m, \quad M_2=\text{ fixed}
\ee
or, stated differently,
\be
\tilde x\to\infty\,,\qquad\gamma\to\infty\,.
\ee
The first condition, $\tilde x\to\infty$ (or, equivalently $\tilde m_1\to \infty$), corresponds to the limit of decoupling of the two mass eigenstates by making one of them very massive. The necessity of the second requirement, \emph{i.e.} taking in addition also $\gamma\to \infty$ (or, equivalently sending $M_1\to\infty$) is less intuitive and requires a more careful explanation. For this let us rewrite the expressions \eqref{rot_angle} and \eqref{masses} for the mixing angle and the mass eigenvalues in terms of $\gamma$, $\tilde x$ leading to
\be\label{tan2theta}
\tan\,2\theta=\frac{2\gamma}{\tilde x^2(1+\gamma^2)+(1-\gamma^2)}\,,
\ee
and
\be\label{masses_v2}
m_{1,2}^2=\frac{1}{2}m^2\left(\tilde x^2+1\pm\frac{\gamma^2(\tilde x^2-1)^2+(\tilde x^2+1)^2}{\gamma^2(\tilde x^2-1)+(\tilde x^2+1)}\cos 2\theta\right)\,.
\ee
Taking the limit of $\tilde x\to \infty$ while keeping $\gamma$ fixed would lead to
\be
\begin{split}
&\lim_{\substack{\tilde x\to \infty,\\\gamma = \text{ fixed}}}\tan 2\theta = 0+\mathcal O\left(\frac{1}{\tilde x^2}\right)\,,\\
 &\lim_{\substack{\tilde x\to \infty,\\\gamma = \text{ fixed}}}m_{1,2}=\frac{1}{2}m^2\left[\tilde x^2+1\pm\left(\tilde x^2+\frac{1-\gamma^2}{1+\gamma^2}+\mathcal O\left(\frac{1}{\tilde x^2}\right)\right)\right]\,.
\end{split}
\ee
In particular, the two eigenmasses in the limit of $\tilde x\to \infty$  up to $\mathcal O(1/\tilde x^2)$ corrections become
\be
m_1^2=m^2\tilde x^2\,,\quad m_2^2=\frac{m^2\gamma^2}{1+\gamma^2}\qquad\Rightarrow\qquad\frac{m_1^2}{m_2^2}=\tilde x^2\left(1+\frac{1}{\gamma^2}\right)\ll 1\,,
\ee
and we have indeed made one of the fields, $h_{\mu\nu}$, much heavier. It is also clear that in order to keep the separation of scales $m_1/M_1=\frac{m}{M_2}\frac{\tilde x}{\gamma}$ fixed, one should set $\gamma\sim\tilde x\to \infty$.

A more physical reason for this is the fact that the $h$ exchange diagrams in the $ff\rightarrow ff$ scattering contribute the following term to the positivity bound:
\be\label{extraterm}
\frac{\text{d}^2}{\text{d}s^2}A_{ff\rightarrow ff} \supset  -\frac{1}{54}\frac{\tilde m_2^4 m_1^2}{M_1^2 m_2^8}(\kappa_2^{(3)})^2 =\left. -\frac{1}{54}\frac{1}{m^2M_2^2}\left(\frac{\tilde x}{\gamma}\right)^2\left(1+\frac{1}{\gamma^2}\right)^4(\kappa_2^{(3)})^2\right|_{\substack{\tilde x\to\infty,\\\gamma=\text{ fixed}}}\,.
\ee
In order to decouple $f$ and $h$ we need this term to vanish. That is not possible to achieve by just setting $\tilde m_1\gg\tilde m_2$ (\emph{i.e.} sending $\tilde x \to \infty$) and keeping $\gamma\sim\mathcal O(1)$ because then this term blows up as $(\tilde x/\gamma)^2\to\infty$.  Instead, the correct limit needs to be
\be\label{decoupling}
\tilde x\to \infty\,,\qquad \gamma\sim \tilde x^p\to \infty \,,\qquad p\geq 1\,,\qquad m,M_2=\text{ fixed},
\ee
thus ensuring that the additional contribution \eqref{extraterm} is (at least) finite and the mass eigenvalues become
\be
m_1=\tilde x m\to\infty\,,\quad m_2=m=\text{ fixed}\,.
\ee

\begin{figure}[t]
    \centering
\includegraphics[height=5.8cm]{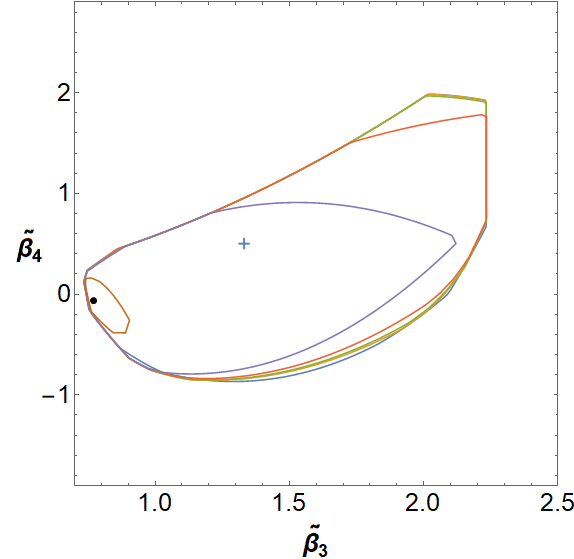}\qquad
\includegraphics[height=5.8cm]{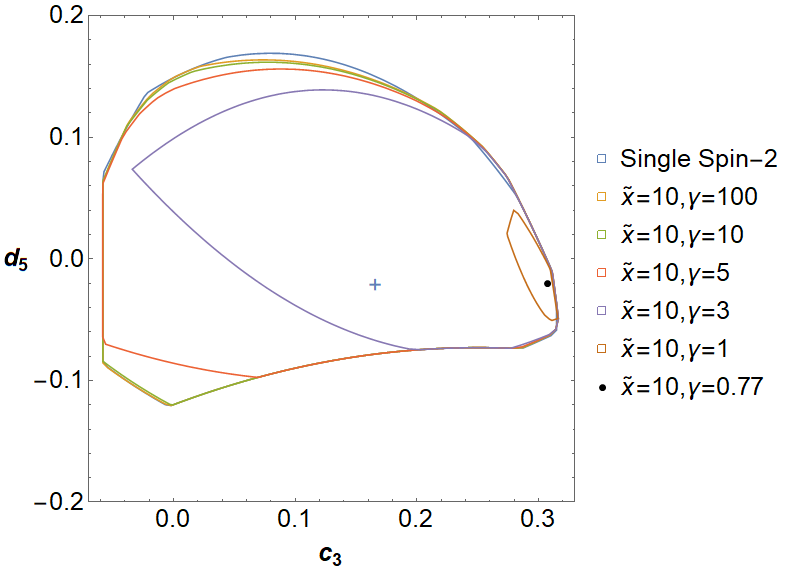}
\caption{The allowed parameter regions in the case of $\tilde x\to \infty$ for different values of $\gamma$. In particular, we set $\tilde x=10$ and $\tilde \alpha_3=\tilde \alpha_4=0$. For large $\gamma$ (as in the limit \eqref{decoupling}) the allowed region reduces to that of a single massive spin-2 field. Decreasing $\gamma$ shrinks the island until it becomes a point at $\gamma_{\rm min}=0.77$. \textbf{Left:} The allowed region in $(\tilde \beta_3,\tilde \beta_4)$ plane. \textbf{Right:} The allowed region in $(c_3,d_5)$ plane, where $\tilde \beta_3=-4c_3+2$ and $\tilde \beta_4=-8d_5-4c_3+1$. }
\label{fig:line cr limit1}
\end{figure}

As a result, in the limit \eqref{decoupling} the heavy field $h$ decouples and the allowed region from $ff\rightarrow ff$ scattering in the $(\tilde \beta_3,\tilde \beta_4)$ plane is the same as that of a single massive spin-2 field. Decreasing $\gamma$ for a fixed mass ratio $\tilde x$ (\emph{i.e.} deviating from the limit \eqref{decoupling}) causes the island to shrink until it becomes a point. For $\tilde x=10$ the minimum value of $\gamma$ was found to be $\gamma_{\rm min}=0.77$. This is shown in Fig.~\ref{fig:line cr limit1}.\\

\begin{figure}[t]
    \centering
\includegraphics[height=5.8cm]{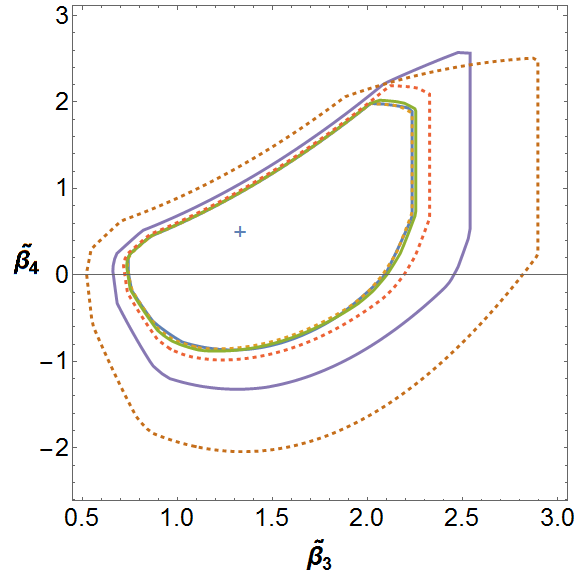}
\qquad
\includegraphics[height=5.8cm]{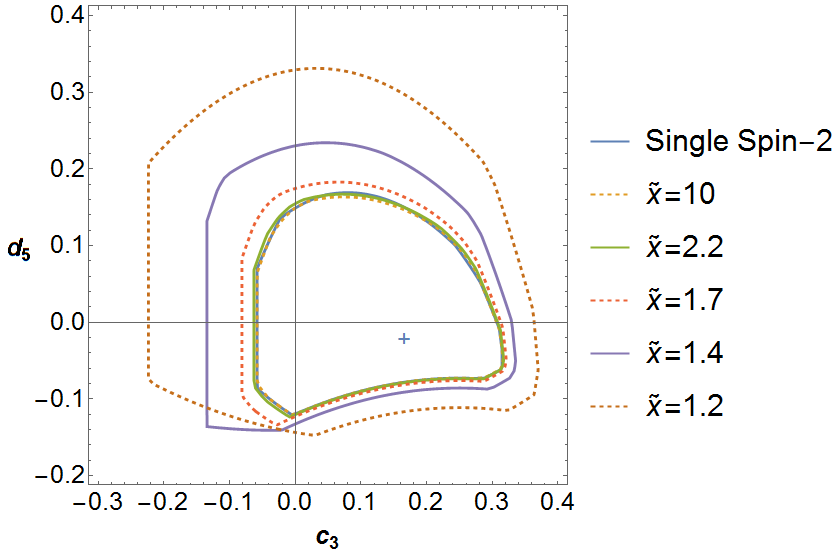}
\caption{The allowed parameter regions in the case of a fixed scaling $\gamma = \tilde x^2$ for different values of $\tilde x$, for $\tilde \alpha_3=\tilde \alpha_4=0$. For large $\tilde x$ (as in the limit \eqref{decoupling}) the allowed region reduces to that of a single massive spin-2 field. Decreasing $\tilde x$ makes the island increase. \textbf{Left:} The allowed region in $(\tilde \beta_3,\tilde \beta_4)$ plane. \textbf{Right:} The allowed region in $(c_3,d_5)$ plane, where $\tilde \beta_3=-4c_3+2$ and $\tilde \beta_4=-8d_5-4c_3+1$. }
\label{fig: singlescale}
\end{figure}

In Fig.~\ref{fig: singlescale} we fix the scaling $\gamma=\tilde x^2$ and show the allowed parameter regions for different values of $\tilde x$. While $\tilde x\to\infty$ and we are well within the limit \eqref{decoupling} we recover the known results for a single massive spin-2 field \cite{Cheung:2016yqr}. Decreasing $\tilde x$ in turn increases the allowed parameter region for $(\tilde\beta_3,\tilde\beta_4)$. The reason for this can be understood in the light of findings of the previous subsection \ref{sec:line_general} in Fig.~\ref{fig:b3b4} --- the tightest bounds on $(\tilde\beta_3,\tilde\beta_4)$ come from the combination of $ff\to ff$ and $fh\to fh$ scattering channels and not from $ff\to ff$ scattering alone.\\

Finally, let us elaborate on the linear dependence on the maximal values of $\tilde x$ on $\gamma$ found in the previous subsection and shown in Fig.~\ref{fig:xt vs gamma}. We saw from \eqref{extraterm} that in the large $\tilde x$ limit, the exchange of the heavy field $h$ introduces an additional \emph{negative} contribution to the $ff\to ff$ positivity bounds that scales as $(\tilde x/\gamma)^2$ for sufficiently large $\gamma$ and $\tilde x$. Moreover in terms of $\tilde x$ and $\gamma$, the other contributions to $A''_{ff\to ff}$ that were omitted in \eqref{extraterm} scale like a `constant' (just the contribution from a single spin-2) with irrelevant corrections as  $\gamma$ and $\tilde x$ are taken to be sufficiently large. It therefore directly follows that the maximal allowed value of $\tilde x$ is linearly related to $\gamma$.

\subsection{Large Mass Gap}

\begin{figure}[h]
    \centering
\includegraphics[height=5.5cm]{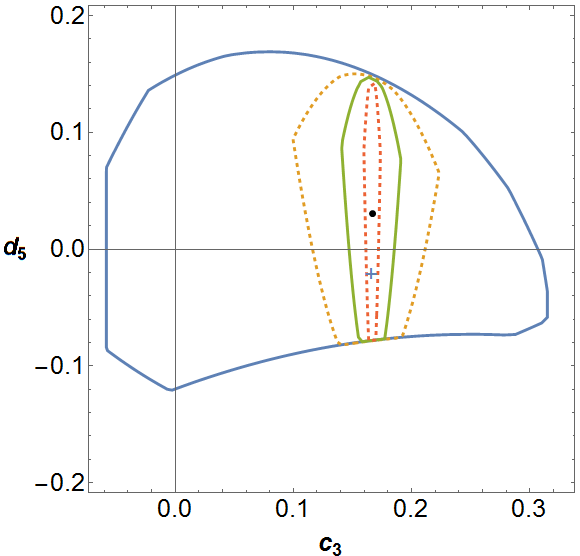}
\qquad
\includegraphics[height=5.5cm]{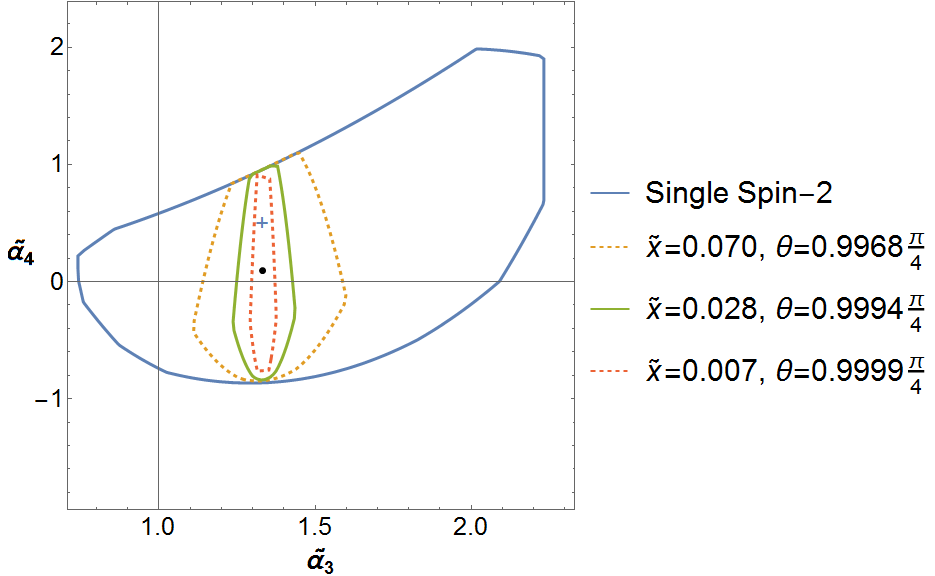}
\caption{The allowed regions in the case of $\gamma=1,\tilde \beta_3=\tilde \beta_4=0$ and different values of $\tilde x\ll 1$. Without knowing about the heavy spin-2 state we would have concluded that the allowed region is the single spin-2 region given by the blue contour. However by using this knowledge and subtracting the heavy pole in the application of positivity bounds we find that the actual region is much smaller. \textbf{Left:} The allowed region in the $(c_3,d_5)$ plane, where $\tilde \alpha_3=-4c_3+2$ and $\tilde \alpha_4=-8d_5-4c_3+1$.  \textbf{Right:} The allowed region in $(\tilde \alpha_3,\tilde \alpha_4)$ plane. }
\label{fig:line cr gamma1}
\end{figure}

Naively, taking one of the masses to be very large should make the scattering amplitudes of the light field to be the same as those of a single massive spin-2 field. One possibility of doing this was discussed in the previous subsection. There is however yet another way of achieving this in the special case when $\gamma=1$. Indeed, in the limit when $\tilde x\to 0$, $\gamma=1$ we see from \eqref{tan2theta} that $\tan 2\theta = 1/\tilde x^2$ and thus $\cos 2\theta= \tilde x^2+\mathcal O (\tilde x^4)$. The physical masses given in \eqref{masses_v2} then become:
\be
\left.\lim_{\substack{\tilde x\to 0}}m_{1,2}^2\right|_{\gamma=1}=\frac{1}{2}m^2\left(1\pm 1 +\mathcal O (\tilde x^2)\right)\,
\ee
leading to $m_1\gg m_2$ thus making the field $h$ much heavier than $f$.
Then the region allowed by positivity bounds imposed on the $ff \to ff$ scattering amplitude might be expected to be the same as for massive gravity. However, we find that this is not the case. In fact the allowed region is much smaller in this limit. The reason for that is the fact that we have subtracted the massive pole coming from the heavy particle from the scattering amplitude as in \eqref{pos}. In other words, we have explicitly used the fact that we know that such a heavy state exists in our theory leading to the additional contribution to the $ff\to ff$ scattering amplitude given in the term \eqref{extraterm}. In this limit the heavy pole is not seen in the EFT, however just by using our knowledge of the spectrum we can significantly reduce the allowed parameter space. This is illustrated in Fig.~\ref{fig:line cr gamma1}, where we plot the allowed regions from the $ff\rightarrow ff$ scattering for small values of $\tilde x$ and $\gamma=1$.

\section{Discussion}\label{sec:conclusions}

In this article we have imposed the now familiar forward limit positivity bounds which arise from the assumption that the EFT admits a local and causal, Lorentz invariant unitary UV completion, to the case of two massive spin-2 fields. This extends the now well developed work on the single massive spin-2 EFT. We have performed this analysis for both line theories, in which interactions between metrics are only pairwise, and cycle theories where not all interactions can be considered as pairwise. We find that the positivity bounds on the scattering amplitudes impose particularly strong constraints on the cubic couplings. For both types of EFTs we show that there is in general a compact region in the coupling parameter space which is consistent with the positivity bounds. This is consistent with earlier work, in particular \cite{Cheung:2016yqr}, however we find that in general the existence of a second spin-2 field imposes stronger constraints even on the self-interactions than implied by single field positivity bounds. For instance, increasing the magnitude of the mixed cubic couplings inevitably shrinks the size of the allowed self-interactions (Cheung-Remmen) parameter island to zero. Hence we infer quite generally that the cubic couplings have a maximum. In the case of cycle theories we also find that they are necessarily positive. If the cubic terms are taken to be zero, then we find that the only mixed quartic interaction that contributes to the positivity bounds must also be zero. \\

Focussing in on the special case of the $\Lambda_3$ theory cycle theory, we find that if the cubic interactions are chosen with the scales implied by the $\Lambda_3$ theory rather than the $\Lambda_{7/2}$ theory, then the quartic interactions are forced to lie in a range in which they also have the appropriate scaling consistent with the $\Lambda_3$ theory. This very much parallels earlier work in which both forward and non-forward limit positivity bounds were applied to the case of a single massive spin-2 field, and it was found that these bounds tend to force the `ghost-free massive gravity' tunings necessary to raise the cutoff of the EFT from $\Lambda_5$ to $\Lambda_3$ \cite{deRham:2018qqo}. This seems to suggest a general theory in which positivity bounds impose or are at least self consistent with technically natural tunings in the EFT Lagrangian which lead to a higher cutoff. It would be interesting to explore this phenomenon in more general contexts. \\

Our analysis shows that the region of allowed parameters implied by indefinite polarization positivity bounds are well captured by a small number ($\sim 10$) of well chosen specific indefinite helicity polarizations which we determine. This may prove useful for future investigations. Interestingly these special polarizations are most clearly stated in the SVT basis, used in \cite{Cheung:2016yqr}, rather than the helicity or transversity polarization basis. This seems to arise because the SVT basis has clearer momentum scaling properties. \\

One of the most interesting aspects of our results is that more fields do not necessarily imply more freedom, and most of the time they reduce it. In the present case the allowed volume of parameter space for the self-interactions of one field is only made smaller by the introduction of a second field. We can understand this from the fact that although more fields lead to more interactions in the EFT Lagrangian, this comes at the price of more scattering amplitudes for more choices of distinct polarizations. Remarkable, at least in these highest cutoff EFTs, it is the latter that wins. We also find a phenomenon noted also in \cite{deRham:2018qqo,Bonifacio:2016wcb} that positivity bounds can in certain cases impose coefficients in the effective action to vanish. This occurs when two different choices of scattering polarizations impose positivity bounds in the manner $f>0$ and $-f>0$. \\

In addition to increasing the effective number of constraints, including more fields also implies including more poles in the low-energy EFT. When those are of spin-2 or more, this leads to additional negative contributions in the positivity bounds and substantially reduce the allowed region of parameter space that could admit a standard UV completion. \\

It would be a straightforward exercise to extend the positivity bounds analysis to any number of massive spin-2 particles coupled together, such as the type of models which arise from dimensional deconstruction \cite{deRham:2013awa,ArkaniHamed:2003vb,Schwartz:2003vj,Deffayet:2003zk,Deffayet:2005yn}. In particular, as we have seen in \cite{Alberte:2019lnd}, as long as the hierarchies between masses $m_i$ are smaller relative to the hierarchies between $m_i$ and the associated interaction scales $M_i$, the results from the decoupling limit about the cutoff scale in the EFT will straightforwardly generalize. In building up a theory of many coupled massive spin-2 fields it would be interesting to understand whether the positivity bounds become increasingly more restrictive, and how that affects the resulting theory. For instance in very specific examples of known partial UV completions of multi spin-2 theories such as Kaluza-Klein theory, the spectrum of masses and couplings is then extremely special (see for example \cite{Bonifacio:2019ioc} for a recent discussion), but more generally spin-2 states may arise not as Kaluza-Klein modes (see \cite{Bachas:2017rch,Bachas:2018zmb,Bachas:2019rfq,deRham:2018dqm} for recent discussions) where there appears to be more freedom. Somewhat less trivially these results will have implications for interacting theories of higher spin. Given any massive spin particle with $s>2$, its decoupling limit will include helicity-two, helicity-one and helicity-zero modes which can similarly be described by the SVT modes discussed above, or by the decoupling limit. The bounds we have considered here will apply to these helicity states of higher spin particles, in addition to the new positivity bounds that will arise there from the new higher helicity states.

\bigskip
\noindent{\textbf{Acknowledgments:}}
We would like to thank Scott Melville for useful discussions.
AJT and CdR would like thank the Perimeter Institute for Theoretical Physics for its hospitality during part of this work and for support from the Simons Emmy Noether program.
LA is supported by the European Research Council under the European Union's Seventh Framework Programme (FP7/2007-2013),
ERC Grant agreement ADG 339140.
The work of AJT and CdR is supported by an STFC grant ST/P000762/1. JR is supported by an STFC studentship. CdR thanks the Royal Society for support at ICL through a Wolfson Research Merit Award. CdR is supported by the European Union's Horizon 2020 Research Council grant 724659 MassiveCosmo ERC-2016-COG and by a Simons Foundation award ID 555326 under the Simons Foundation's Origins of the Universe initiative, `\textit{Cosmology Beyond Einstein's Theory}'. AJT thanks the Royal Society for support at ICL through a Wolfson Research Merit Award.

\bigskip

\appendix

\section{Generic Diagonalization of a Line of Interactions}\label{sec:diag}

We start with the canonically normalized variables $\tilde h\mn$ and $\tilde f\mn$ defined in terms of the would-be metrics as in \eqref{metric_def}. Then perturbatively the tensor $\K\mupn$ defined in \eqref{Kmunu} is given by
\begin{align}
\K\mupn&(g^{(1)},g^{(2)})=\frac{\tilde h\mupn}{M_1}-\frac{\tilde f\mupn}{M_2}-\frac 12 \(2\frac{\tilde h^{\mu\alpha}\tilde h_{\alpha\nu}}{M_1^2}+\frac{\tilde f^{\mu\alpha}\tilde h_{\alpha\nu}}{M_1M_2}-3\frac{\tilde h^{\mu\alpha}\tilde f_{\alpha\nu}}{M_1M_2}\)\\
&+\frac14\(
4\frac{\tilde h^3{}\mupn}{M_1^3} +\frac{\tilde f^{\mu\alpha}\tilde h^2_{\alpha\nu}}{M_2M_1^2}+2 \frac{\tilde h^{\mu\alpha}\tilde f_\alpha^\beta \tilde h_{\beta \nu}}{M_2M_1^2}+\frac{\tilde f^2{}^{\mu\alpha}\tilde h_{\alpha\nu}}{M_2^2M_1}-7\frac{\tilde h^2{}^{\mu\alpha}\tilde f_{\alpha\nu}}{M_1^2M_2}
+\frac{\tilde h^{\mu\alpha}\tilde f^2_{\alpha \nu}}{M_1M_2^2}-2 \frac{\tilde f^{\mu \alpha}\tilde h_\alpha^\beta \tilde f_{\beta_\nu}}{M_1M_2^2}
\)+\cdots\nn\,,
\end{align}
where ellipses include terms quartic and higher order in $\tilde h\mn$ and $\tilde f\mn$ and all indices are raised and lowered with respect to $\eta\mn$. Note that $\K\mn=\eta_{\mu\alpha}\K^\alpha_{\,\nu}$ is {\it not symmetric} but $\tilde \K\mn=g^{(1)}_{\mu\alpha}\K^\alpha_{\,\nu}$ is symmetric as expected. The action \eqref{eq:actionline} at quadratic level then becomes
\ba
\label{eq:line2}
\L^{(2)}_{\rm line}&=&- \tilde h^{\mu \nu }\mathcal{E}^{\alpha\beta}\mn \tilde h\ab- \tilde f^{\mu \nu }\mathcal{E}^{\alpha\beta}\mn \tilde f\ab-\frac 12 \tilde m_1^2 \([\tilde h^2]-[\tilde h]^2\)\\
&-&\frac 12 \frac{\tilde m_2^2}{ M_1^2+ M_2^2} \(\left[\( M_1\tilde f- M_2\tilde h\)^2\right]-\left[ M_1 \tilde f- M_2\tilde h\right]^2\)\,,\nn
\ea
and can be diagonalized by performing the field space rotation $\begin{pmatrix}\tilde h_{\mu\nu}\\\tilde f_{\mu\nu}\end{pmatrix}=\mathcal R^T(\theta)\begin{pmatrix} h_{\mu\nu}\\ f_{\mu\nu}\end{pmatrix}$, with rotation angle $\theta$ given in \eqref{rot_angle}. The resulting quadratic Lagrangian is then
\ba
\L^{(2)}_{\rm line}=- h^{\mu \nu }\mathcal{E}^{\alpha\beta}\mn  h\ab-   f^{\mu \nu }\mathcal{E}^{\alpha\beta}\mn  f\ab-\frac 12  m_1^2 \([ h^2]-[ h]^2\)-\frac 12  m_2^2 \([ f^2]-[f]^2\)\,,
\ea
with the physical eigenmasses $m_{1,2}$ given by \eqref{masses} in section~\ref{sec:main}.

As already presented in the main text, after diagonalization the cubic and quartic kinetic terms in terms of the mass eigenmodes $h_{\mu\nu}$ and $f_{\mu\nu}$ are symbolically of the form:
\ba
\L^{(3)}_{\rm Kin}&=&\frac{1}{M_1}\p^2 (c h + s f)^3+\frac{1}{ M_2}\p^2 (c f - s h)^3\,,\\
\L^{(4)}_{\rm Kin}&=&\frac{1}{M_1^2}\p^2 (c h + s f)^4+\frac{1}{ M_2^2}\p^2 (c f - s h)^4\,,
\ea
where in here and in what follows we use the notation $c\equiv \C$ and $s\equiv \S$ and
where $\p^2 h^3 $ and  $\p^2 h^4$) representation the standard Einstein-Hilbert term at cubic and quartic order for metric perturbations defined as in \eqref{metric_def}. The potential terms in turn  take the form
\ba
\L^{(3)}_{\rm mass}&=&\frac{\tilde m_2^2}{4 M_1}\ \sum_{n=0}^3
\kappa^{(3)}_{n} \E \E h^{3-n} f^{n} \\
\L^{(4)}_{\rm mass}&=&\frac{\tilde m_2^2}{4 M_1^2}\ \sum_{n=0}^4
\kappa^{(4)}_{n} \E \E h^{4-n} f^{n}+\frac{ \tilde m_2^2}{ 4(M_1^2+ M_2^2)}\([f\cdot h\cdot f\cdot h]-[f^2\cdot h^2]\)
\ea
where the expressions for the coefficients $\kappa^{(3,4)}_n$ are given by
\ba\label{cubic_kappas}
\kappa^{(3)}_n&=&
\frac{2}{ M_1^2+ M_2^2}\(\begin{array}{c}
c (c M_2+ s  M_1)^2 \\
s^3  M_1^2+4 c s^2  M_1  M_2+ c^2 s (-2  M_1^2+3 M_2^2)-2 c^3  M_1  M_2
\\
c^3  M_1^2-4 c^2 s  M_1  M_2+ c s^2 (-2  M_1^2+3 M_2^2)+2 s^3  M_1  M_2\\
s (s M_2-c  M_1)^2
\end{array}
\)\\
&+&\frac{\tilde m_1^2}{\tilde m_2^2}\(\begin{array}{c}
c^3 \\ 3 c^2 s \\ 3 c s^2 \\ s^3
\end{array}
\)\tilde \alpha_3
+
\frac{1}{ M_2 ( M_1^2+ M_2^2)}\(\begin{array}{c}
\(c M_2+s  M_1\)^3 \\
3 \(c M_2+s  M_1\)^2 \(s M_2-c  M_1\) \\
3 \(c M_2+s  M_1\) \(s M_2-c  M_1\)^2 \\
\(s M_2-c  M_1\)^3
\end{array}
\)\tilde \beta_3\nn\,,
\ea
and
\ba\label{quartic_kappas}
\kappa^{(4)}_n&=&
\frac{1}{M_1^2+M_2^2}\(\begin{array}{c}
c^2 (c M_2+ s  M_1)^2 \\
-2c\(c M_2+s  M_1\)\(c^2 M_1-2 c s  M_2-s^2 M_1\)\\
\(c^2 M_1-2 c s  M_2-s^2  M_1\)^2 -2 c s \(s M_2-c  M_1\)\(c M_2+s  M_1\)
\\
-2s\(s M_2-c M_1\)\(c^2 M_1-2 c s  M_2-s^2  M_1\)\\
s^2 (s M_2-c  M_1)^2
\end{array}
\)
\\
&+&\frac{\tilde m_1^2}{\tilde m_2^2}\(\begin{array}{c}
c^4 \\ 4 c^3 s \\ 6 c^2 s^2 \\ 4 c s^3 \\ s^4
\end{array}
\)\tilde \alpha_4
+
\frac{1}{ M_2 ( M_1^2+ M_2^2)}\(\begin{array}{c}
c\(c M_2+s M_1\)^3 \\
\(c M_2+s  M_1\)^2 \(s^2  M_1 +4 c s  M_2 -3 c^2  M_1\) \\
3\(c M_2+s  M_1\)\(s M_2-c  M_1\)\(c^2 M_1 -2 c s  M_2 - s^2 M_1\)\\
\(s M_2-c  M_1\)^2 \(-c^2  M_1 +4 c s M_2 +3 s^2  M_1\)\\
s\(s M_2-c  M_1\)^3
\end{array}
\)\tilde \beta_3\nn\\
&+&
\frac{1}{ M_2^2 ( M_1^2+ M_2^2)}\(\begin{array}{c}
\(c M_2+s  M_1\)^4 \\
4 \(c M_2+s  M_1\)^3 \(s M_2-c  M_1\) \\
6 \(c M_2+s  M_1\)^2 \(s M_2-c  M_1\)^2 \\
4 \(c M_2+s M_1\) \(s M_2-c  M_1\)^3 \\
\(sM_2-c  M_1\)^4
\end{array}
\)\tilde \beta_4\nn\,.
\ea

\section{Conventions}\label{sec:conventions}

\subsection{Mandelstam Variables}\label{mand vars}
For $2-2$ scattering amplitudes we use the Mandelstam variables defined as
\begin{align}
    s=-(k_1+k_2)^2\,, \quad t=-(k_1-k_3)^2\,, \quad u=-(k_1-k_4)^2\,,
\end{align}
where $k^\mu_i=(E_i,\textbf{p}_i)$ for $i=1,2,3,4$. The on-shell condition for each of the particles reads $k_\mu k^\mu = m_i^2$ as usual. The Mandelstam variables satisfy the following relation:
\begin{equation}
    s+t+u=\sum_{i}^4m^2_{i}\,,
\end{equation}
and, hence, one can express the amplitude in terms of two independent variables. For an elastic scattering process with $m_1=m_3$ and $m_2=m_4$, a convenient choice of variables is to define
\begin{equation}
   \mathcal{S}=(s-(m_1-m_2)^2)(s-(m_1+m_2)^2)\,.
\end{equation}
 In the center of mass frame we can write the energies and the three-momenta as:
\begin{align}
    \textbf{p}_1^2 &= \frac{1}{4s}[s-(m_1+m_2)^2][s-(m_1-m_2)^2]\,,\\
    \textbf{p}_3^2 &= \frac{1}{4s}[s-(m_3+m_4)^2][s-(m_3-m_4)^2]\,,\\
     E_1 &= \frac{1}{2\sqrt{s}}(s+m_1^2-m_2^2)\,, \quad E_2=\frac{1}{2\sqrt{s}}(s+m_2^2-m_1^2)\,, \\
    E_3 &= \frac{1}{2\sqrt{s}}(s+m_3^2-m_4^2)\,, \quad E_4=\frac{1}{2\sqrt{s}}(s+m_4^2-m_3^2)\,.
\end{align}
For $m_1=m_3$, $m_2=m_4$ we can simplify these relations as
\begin{equation}
    \textbf{p}_1^2=\textbf{p}_3^2\equiv p^2=\frac{1}{4s}\mathcal S
\end{equation}
and
\begin{equation}
     E_1 =E_3= \frac{1}{2\sqrt{s}}(s+m_1^2-m_2^2) \,, \qquad  E_2=E_4 = \frac{1}{2\sqrt{s}}(s-m_1^2+m_2^2)\,.
\end{equation}
We write the four-momenta in spherical coordinates with $\phi=0$ as
\begin{equation}\label{scat_angle}
    p^{\mu}_i=(E_i,p\sin{\theta_i},0,p\cos{\theta_i})\,,
\end{equation}
where $E_i$ and $p$ are the energy and the absolute value of the spatial momenta in the center of mass frame. For the $2-2$ scattering, we define the scattering angles for each particle as $\theta_1 = 0,~\theta_2 = \pi,~\theta_3=\theta,~\theta_4=\pi+\theta$. Finally, we express $t$ in terms of the scattering angle, $\theta$, as
\begin{equation*}
\cos{\theta} = 1 + \frac{2ts}{\mathcal S}\,.
\end{equation*}

\subsection{Polarization Tensors}\label{app:pol tens}
To construct  the SVT basis of polarization tensors we first decompose the space orthogonal to $k_\mu$ in terms of three polarization basis vectors, $\epsilon_\mu^i$, satisfying $p^\mu\epsilon^i_\mu=0$. In a frame where $\theta=0$ and $p_\mu=(E,0,0,p)$ such basis vectors are defined as:
\begin{align}\label{svt vec}
    \epsilon_{\mu}^1 &= (0,1,0,0)\,, \\
    \epsilon_{\mu}^2 &= (0,0,1,0)\,, \\
     \epsilon_{\mu}^3 &\equiv\epsilon_{\mu}^S = \frac{1}{m}(p,0,0,E)\,.\label{svt vec1}
\end{align}
Using the same convention as in \cite{Cheung:2016yqr} we define the polarization tensors in this basis so as to satisfy
\be
p^\mu\epsilon^i_{\mu\nu}=\epsilon_\mu^i\,^\mu=0\,.
\ee
These can be obtained as
\begin{align}
     \epsilon_{\mu\nu}^{T1} &= \frac{1}{\sqrt{2}}(\epsilon_{\mu}^1\epsilon_{\nu}^1-\epsilon_{\mu}^2\epsilon_{\nu}^2)\,, \label{svt T1}\\
     \epsilon_{\mu\nu}^{T2} &= \frac{1}{\sqrt{2}}(\epsilon_{\mu}^1\epsilon_{\nu}^2+\epsilon_{\mu}^2\epsilon_{\nu}^1)\,, \label{svt T2}\\
     \epsilon_{\mu\nu}^{V1} &= \frac{i}{\sqrt{2}}(\epsilon_{\mu}^1\epsilon_{\nu}^3+\epsilon_{\mu}^3\epsilon_{\nu}^1)\,, \\
      \epsilon_{\mu\nu}^{V2} &= \frac{i}{\sqrt{2}}(\epsilon_{\mu}^2\epsilon_{\nu}^3+\epsilon_{\mu}^3\epsilon_{\nu}^2)\,, \\
       \epsilon_{\mu\nu}^{S} &= \sqrt{\frac{3}{2}}\bigg(\epsilon_{\mu}^3\epsilon_{\nu}^3-\frac{1}{3}\bigg(\eta_{\mu\nu}+\frac{p_{\mu}p_{\nu}}{m^2}\bigg)\bigg)\,.\label{last_pol}
\end{align}

\pagebreak
\section{Scattering Amplitudes in SVT Basis and Special Polarizations}\label{boundsdef}
Here we give the full expressions for the scattering amplitudes for the cycle theories computed for the indefinite helicity states, as defined in \eqref{indefpolhel}, as well as the special choice of indefinite helicity polarizations for both cycle and line theories that reproduce most efficiently the plots in the main text. We consider the $hf\to hf$ and $hh\to hh$ scattering processes ($ff\to ff$ is straightforwardly determined by exchanging parameters). We also present the precise choices of polarizations of the ingoing and outgoing particles leading to the strongest constraints shown as the parameter islands in Figs.~\ref{fig:c3d5x.5} and \ref{fig:c3d5x.1}. To find the allowed region from indefinite scattering we used a similar but not identical method to \cite{Cheung:2016yqr}. We picked points in the parameter space and minimized the function \ref{pos} with respect to $\alpha$'s and $\beta$'s in \eqref{indefpolhel}. If the point in the parameter space was not allowed that gave the inequality excluding it. We then plotted the obtained inequalities in 2d space and repeated picking the points until we obtain a compact region. Then we repeated for the points in this region until it was not possible to exclude any more points. The net result is a more efficient means to obtain the island in parameter space from only a handful of well chosen polarization states.

\begin{table}[H]\normalsize
    \centering
    \begin{tabular}{|c|c|c|c|c|}
    \hline
         $\alpha_{T1}$ $\beta_{T1}$&$\alpha_{T2}$ $\beta_{T2}$ & $\alpha_{V1}$ $\beta_{V1}$ & $\alpha_{V2}$ $\beta_{V2}$ & $\alpha_{S}$ $\beta_{S}$ \\
                  \hhline{|=|=|=|=|=|}
         $0.650,0.650$ &$0.512,-0.512$& $0.454,-0.454$ &$0.157,0.157$&$-0.292,-0.292$\\
         \hline
         $-0.909,0.909$ &$0.349,0.349$& $0,0$ &$0,0$&$0.227,-0.227$\\
         \hline
         $0.568,0.568$ &$-0.493,0.493$& $0,0$ &$0,0$&$-0.659,0.659$\\
         \hline
         $0.358,0.358$ &$0.907,-0.907$& $0.125,-0.125$ &$-0.184,-0.184$&$-0.005,-0.005$\\
         \hline
         $-0.195,-0.195$ &$0.975,-0.975$& $0.065,-0.065$ &$0.080,0.080$&$0.0004,0.0004$\\
         \hline
         $0.756,-0.756$ &$0.487,0.487$& $0,0$ &$0,0$&$0.438,0.438$\\
         \hline
         $-0.223,-0.223$ &$0.449,-0.449$& $-0.642,0.642$ &$0.398,0.398$&$0.421,0.421$\\
         \hline
         $0,0$ &$0,0$& $0.590,-0.590$ &$0,0$&$0.808,0.808$\\
  \hline\hline
         $0,0$ &$0,0$& $-0.999,-0.041$ &$-0.041,-0.999$&$0,0$\\
         \hline
         $0.701,0.690$ &$-0.411,0.404$& $0,0$ &$0,0$&$-0.583,-0.601$\\
         \hline
         $0.677,0.795$ &$0.538,-0.232$& $0.406,-0.637$ &$0.084,0.131$&$-0.534,-0.483$\\
         \hline
         $0.116,-0.001$ &$-0.059,-0.001$& $-0.174,-0.200$ &$-0.724,0.833$&$0.655,-0.516$\\

    \hline
    \end{tabular}
    \caption{ Special configurations of polarizations for $hf\rightarrow hf$ scatterings in SVT basis which give strong constraints on the couplings. These are determined by minimizing the $\alpha$'s and $\beta$'s for guessed Lagrangian parameters. This procedure is repeated until a compact parameter region is obtained. In the case of line theories, the last four configurations of polarizations are needed, in addition to the first seven, to give strong constraints on the couplings.}
    \label{tab:2}
\end{table}

\subsection{$hf\rightarrow hf$ Scattering}\label{hfhfSVT}
\begin{equation}\label{indef bound}
\begin{split}
f=&\frac{g_*^2}{\gamma \Lambda_2^4}\Bigg(\Bigg.\frac{ (4 \alpha_{S}^2+3 \alpha_{V1}^2+3 \alpha_{V2}^2) (4 \beta_{S}^2+3 \beta_{V1}^2+3 \beta_{V2}^2)}{12  x^2}( \kappa^{(2)}_{3} c_{1}+ \kappa^{(1)}_{3} c_{2})\\
&+\frac{c_{2} (4 \beta_{S}^2+3 \beta_{V1}^2+3 \beta_{V2}^2) (2 \alpha_{S}^2+6 \alpha_{T1}^2+6 \alpha_{T2}^2+3 \alpha_{V1}^2+3 \alpha_{V2}^2)}{18  x^2}\\
&+\frac{c_{1} (4 \alpha_{S}^2+3 \alpha_{V1}^2+3 \alpha_{V2}^2) (2 \beta_{S}^2+6 \beta_{T1}^2+6 \beta_{T2}^2+3 \beta_{V1}^2+3 \beta_{V2}^2)}{18  x^2}\\
-&\Bigg(\Bigg.\alpha _{S}^2 (8 \beta _{S}^2+6 (\beta _{V1}^2+\beta _{V2}^2))-6 \alpha _{S}\bigg(4\beta _{S}( \alpha _{T1}\beta _{T1}- \alpha _{T2}\beta _{T2})+\sqrt{3} \alpha _{T1} (\beta _{V2}^2-\beta _{V1}^2)+2 \sqrt{3} \alpha _{T2} \beta _{V1} \beta _{V2}\bigg)\\+&6 \beta _{S}^2 (\alpha _{V1}^2+\alpha _{V2}^2)+6 \sqrt{3} \beta _{S} ((\alpha _{V1}^2 -\alpha _{V2}^2) \beta _{T1}-2 \alpha _{V1} \alpha _{V2} \beta _{T2})+9 (\alpha _{V1} \beta _{V2}+\alpha _{V2} \beta _{V1})^2\Bigg. \Bigg)\frac{\lambda}{18  x^2}\\
+\Bigg(\Bigg.&4 \alpha _{S}^2 \beta _{S}^2 x^5+12 \alpha _{S} \beta _{S} x^4 (\alpha _{V1} \beta _{V1}-\alpha _{V2} \beta _{V2})+x^3 \bigg(-32 \alpha _{S}^2 \beta _{S}^2+48 \alpha _{S} \beta _{S} (\alpha _{T2} \beta _{T2}-\alpha _{T1} \beta _{T1})\\+&18 x^2 (6 \alpha _{S} \beta _{S} (\alpha _{V1} \beta _{V1}-\alpha _{V2} \beta _{V2})-\sqrt{3} \beta _{S} (\beta _{V1} (\alpha _{T1} \alpha _{V1}+\alpha _{T2} \alpha _{V2})-\beta _{V2} (\alpha _{T2} \alpha _{V1}-\alpha _{T1} \alpha _{V2}))\\+&6 \alpha _{S} (\beta _{S} (\alpha _{V1} \beta _{V1}-\alpha _{V2} \beta _{V2})-3 \sqrt{3} (\beta _{V1} (\alpha _{V1} \beta _{T1}-\alpha _{V2} \beta _{T2})+\beta _{V2} (\alpha _{V1} \beta _{T2}+\alpha _{V2} \beta _{T1})))\\+&9 (\alpha _{V1}^2 (\beta _{S}^2-3 \beta _{V1}^2)+6 \alpha _{V1} \alpha _{V2} \beta _{V1} \beta _{V2}+\alpha _{V2}^2 (\beta _{S}^2-3 \beta _{V2}^2))\big)\\+&4 (\alpha _{T1} \beta _{T1}-\alpha _{T2} \beta _{T2}) (\alpha _{V1} \beta _{V1}-\alpha _{V2} \beta _{V2})\bigg)-x \big(9 \big(\alpha _{S}^2 (\beta _{V1}^2+\beta _{V2}^2)\\+&2 \sqrt{3} \alpha _{S} (\alpha _{T1} (\beta _{V2}^2-\beta _{V1}^2)+2 \alpha _{T2} \beta _{V1} \beta _{V2})+5 (\alpha _{V1} \beta _{V1}-\alpha _{V2} \beta _{V2})^2\big)+2 \beta _{S}^2 (16 \alpha _{S}^2+9 (\alpha _{V1}^2+\alpha _{V2}^2))\\+&6 \beta _{S} (16 \alpha _{S} (\alpha _{T1} \beta _{T1}-\alpha _{T2} \beta _{T2})+3 \sqrt{3} (\beta _{T1} (\alpha _{V2}^2-\alpha _{V1}^2)+2 \alpha _{V1} \alpha _{V2} \beta _{T2}))\big)\Bigg. \Bigg)\frac{c_{2}^2}{54  x^3}\gamma\\
+\Bigg(\Bigg.&x^5 \big(6 \alpha _{S} \beta _{S} (\alpha _{V1} \beta _{V1}-\alpha _{V2} \beta _{V2})-18 \sqrt{3} \beta _{S} (\beta _{V1} (\alpha _{T1} \alpha _{V1}+\alpha _{T2} \alpha _{V2})+\beta _{V2} (\alpha _{T1} \alpha _{V2}-\alpha _{T2} \alpha _{V1}))\big)\\+&x^4 \bigg(-32 \alpha _{S}^2 \beta _{S}^2-18 \alpha _{S}^2 (\beta
_{V1}^2+\beta _{V2}^2)+6 \alpha _{S} \bigg(-16 \alpha
_{T1} \beta _{S} \beta _{T1}+3 \sqrt{3} \alpha _{T1}(\beta _{V1}^2-\beta _{V2}^2)\\+&2 \alpha _{T2} (8\beta _{S} \beta _{T2}-3 \sqrt{3} \beta _{V1} \beta_{V2})\bigg)-9 \bigg(\alpha _{V1}^2 (\beta _{S}^2--2\sqrt{3} \beta _{S} \beta _{T1}+5 \beta _{V1}^2)+2 \alpha_{V1} \alpha _{V2} (2 \sqrt{3} \beta _{S} \beta _{T2}-5\beta _{V1} \beta _{V2})\\+&\alpha _{V2}^2 (\beta_{S}^2+2 \sqrt{3} \beta _{S} \beta _{T1}+5 \beta_{V2}^2)\bigg)\bigg)\\+&18 x^3 \bigg(\alpha _{S} (6 \beta _{S} (\alpha _{V1} \beta _{V1}-\alpha _{V2} \beta _{V2})-\sqrt{3} (\beta _{V1} (\alpha _{V1} \beta _{T1}-\alpha _{V2} \beta _{T2})+\beta _{V2} (\alpha _{V1} \beta _{T2}+\alpha _{V2} \beta _{T1})))\\+&4 \alpha _{T1} \beta _{T1} (\alpha _{V1} \beta _{V1}-\alpha _{V2} \beta _{V2})-4 \alpha _{T2} \beta _{T2} (\alpha _{V1} \beta _{V1}-\alpha _{V2} \beta _{V2})\bigg)\\+&x^2 \bigg(-32 \alpha _{S}^2 \beta _{S}^2+9 \alpha _{S}^2 (\beta _{V1}^2+\beta _{V2}^2)+48 \alpha _{S} \beta _{S} (\alpha _{T2} \beta _{T2}-\alpha _{T1} \beta _{T1})-27 (\alpha _{V1} \beta _{V1}-\alpha _{V2} \beta _{V2})^2\bigg)\\+&12x \alpha _{S} \beta _{S}  (\alpha _{V1} \beta _{V1}-\alpha _{V2} \beta _{V2})+4 \alpha _{S}^2 \beta _{S}^2\Bigg. \Bigg)\frac{c_{1}^2}{54  x^8}\frac{1}{\gamma}\Bigg. \Bigg)>0.
\end{split}
\end{equation}

\subsection{$hh\rightarrow hh$ Scattering}\label{hhhhSVT}

\allowdisplaybreaks

\begin{align}
\begin{autobreak}\label{Island}

f=
-\frac{g_*^2}{432\gamma^2\Lambda_2^4 x^4}\bigg( 4\bigg(2 (4 \alpha_{S} \beta_{S}-3 \alpha_{V1} \beta_{V1}+3 \alpha_{V2} \beta_{V2}) (8 \alpha_{S}\beta_{S}
+12 \alpha_{T1} \beta_{T1}
-12 \alpha_{T2} \beta_{T2}
-9 \alpha_{V1} \beta_{V1}
+9 \alpha_{V2} \beta_{V2}) x^6
+(-4 (16  \beta_{S}^2+9 (\beta_{V1}^2
+\beta_{V2}^2)) \alpha_{S}^2
+6 (8 \alpha_{T1} \beta_{S} \beta_{T1}
+3 \sqrt{3}\alpha_{V1} \beta_{V1} \beta_{T1}
+3 \sqrt{3} \alpha_{V2} \beta_{V2} \beta_{T1}
-8 \alpha_{T2} \beta_{S} \beta_{T2}
-4\alpha_{V1} \beta_{S} \beta_{V1}
-3 \sqrt{3} \alpha_{V2} \beta_{T2} \beta_{V1}
+4 \alpha_{V2} \beta_{S} \beta_{V2}
+3 \sqrt{3}
   \alpha_{V1} \beta_{T2} \beta_{V2}
+6 \sqrt{3} \alpha_{T2} \beta_{V1} \beta_{V2}
+3 \sqrt{3} \alpha_{T1} (\beta_{V2}^2
-\beta_{V1}^2)) \alpha_{S}
+9 (-(\beta_{V1}^2
+3 \beta_{V2}^2) \alpha_{V1}^2
-4 \alpha_{V2} \beta_{V1}
   \beta_{V2} \alpha_{V1}
-4 (\alpha_{V1}^2
+\alpha_{V2}^2) \beta_{S}^2
-\alpha_{V2}^2 (3 \beta_{V1}^2
+\beta _{V2}^2)
+2 \sqrt{3} \beta_{S} (
-\beta_{T1} \alpha_{V1}^2
+(2 \alpha_{V2} \beta_{T2}
+\alpha_{T1} \beta_{V1}
-\alpha_{T2}
   \beta_{V2}) \alpha_{V1}
+\alpha_{V2} (\alpha_{V2} \beta_{T1}
+\alpha_{T2} \beta_{V1}
+\alpha_{T1}\beta_{V2})))) x^4
+((8 \beta_{S}^2
+9 (\beta_{V1}^2
+\beta_{V2}^2)) \alpha_{S}^2
+18 \beta_{S} (\alpha_{V1}
   \beta_{V1}
-\alpha_{V2} \beta_{V2}) \alpha_{S}
+9 (\alpha_{V1}^2
+\alpha_{V2}^2) \beta_{S}^2) x^2+4 \alpha_{S}^2 \beta_{S}^2\bigg) \frac{c_1^2}{x^4}
+27 x^2 \bigg((4 \beta_{S}^2-9 (\beta_{V1}^2+\beta_{V2}^2)) \alpha_{S}^2+6 (8 \alpha_{T1} \beta_{S} \beta_{T1}+\sqrt{3} \alpha_{V1}\beta_{V1} \beta_{T1}+\sqrt{3} \alpha_{V2} \beta_{V2} \beta_{T1}
-8 \alpha_{T2} \beta_{S} \beta_{T2}
-7 \alpha_{V1} \beta_{S} \beta_{V1}
-\sqrt{3} \alpha_{V2} \beta_{T2} \beta_{V1}
+7 \alpha_{V2}\beta_{S} \beta_{V2}
+\sqrt{3} \alpha_{V1} \beta_{T2} \beta_{V2}
+2 \sqrt{3} \alpha_{T2} \beta_{V1} \beta_{V2}
+\sqrt{3} \alpha_{T1} (\beta_{V2}^2
-\beta_{V1}^2)) \alpha_{S}
+3 (-3 (\alpha_{V1}^2
+\alpha_{V2}^2) \beta_{S}^2
+2 \sqrt{3} (-\beta_{T1} \alpha_{V1}^2
+(2 \alpha_{V2} \beta_{T2}
+\alpha_{T1} \beta_{V1}
-\alpha_{T2} \beta_{V2}) \alpha_{V1}
+\alpha_{V2} (\alpha_{V2} \beta_{T1}
+\alpha_{T2} \beta_{V1}
+\alpha_{T1} \beta_{V2})) \beta_{S}
+5 \alpha_{V1}^2\beta_{V1}^2
-3 \alpha_{V2}^2 \beta_{V1}^2
-3 \alpha_{V1}^2 \beta_{V2}^2
+5 \alpha_{V2}^2 \beta_{V2}^2
-8 \alpha_{T1} \alpha_{V1} \beta_{T1} \beta_{V1}
+8 \alpha_{T2} \alpha_{V1} \beta_{T2} \beta_{V1}
+8 \alpha_{T1} \alpha_{V2} \beta_{T1}\beta_{V2}
-8 \alpha_{T2} \alpha_{V2} \beta_{T2} \beta_{V2}
-16 \alpha_{V1} \alpha_{V2} \beta_{V1} \beta_{V2})\bigg) \kappa_{3}^2
+18 x^2 \bigg(-(64 \beta_{S}^2+24 \beta_{T1}^2+24 \beta_{T2}^2+9 (\beta_{V1}^2+\beta_{V2}^2)) \alpha_{S}^2-6 (16 \alpha_{T1} \beta_{S}\beta_{T1}+2 \alpha_{T2} (\sqrt{3} \beta_{V1}
   \beta_{V2}
-8 \beta_{S} \beta_{T2})
+\sqrt{3} \alpha_{T1} (\beta_{V2}^2
-\beta_{V1}^2)
+5 (-5 \alpha_{V1}\beta
   _{S} \beta_{V1}
+\alpha_{V2} (
-\sqrt{3} \beta_{T2} \beta_{V1}
+5 \beta_{S} \beta_{V2}
+\sqrt{3} \beta_{T1}\beta_{V2})
+\sqrt{3} \alpha_{V1} (\beta_{T1} \beta_{V1}
+\beta_{T2} \beta_{V2}))) \alpha_{S}
-3 (8 \alpha_{T1}^2
+8 \alpha_{T2}^2
+3 (\alpha_{V1}^2
+\alpha_{V2}^2)) \beta_{S}^2
+6 \sqrt{3} \beta_{S} (\beta_{T1} \alpha_{V1}^2
+(-2 \alpha_{V2} \beta_{T2}
-5 \alpha_{T1} \beta_{V1}
+5 \alpha_{T2} \beta_{V2}) \alpha_{V1}
-\alpha_{V2} (\alpha_{V2}\beta_{T1}
+5 \alpha_{T2} \beta_{V1}
+5 \alpha_{T1} \beta_{V2}))
+9 (-(2 \beta_{T1}^2
+2 \beta_{T2}^2
+7 \beta_{V1}^2
+\beta _{V2}^2) \alpha_{V1}^2
+4 \beta_{V1} (2 \alpha_{T1} \beta_{T1}
-2 \alpha_{T2} \beta_{T2}
+3 \alpha_{V2} \beta_{V2}) \alpha_{V1}
+8 \alpha_{V2} (\alpha_{T2} \beta_{T2}
-\alpha_{T1} \beta_{T1}) \beta_{V2}
-2 (\alpha_{T1}^2
+\alpha_{T2}^2)
   (\beta_{V1}^2
+\beta_{V2}^2)
-\alpha_{V2}^2 (2 \beta_{T1}^2
+2 \beta_{T2}^2
+\beta_{V1}^2
+7 \beta_{V2}^2))\bigg)\kappa_3
+144 x^2 \bigg((8 \beta_{S}^2+6 (\beta_{V1}^2+\beta_{V2}^2)) \alpha_{S}^2-6 (4 \alpha_{T1} \beta_{S}\beta_{T1}-4 \alpha_{T2} \beta_{S} \beta_{T2}
+2 \sqrt{3} \alpha_{T2} \beta_{V1} \beta_{V2}
+\sqrt{3} \alpha_{T1} (\beta_{V2}^2
-\beta_{V1}^2)) \alpha_{S}
+6 (\alpha_{V1}^2
+\alpha_{V2}^2) \beta_{S}^2
+9 (\alpha_{V2}\beta_{V1}
+\alpha_{V1} \beta_{V2})^2
+6 \sqrt{3} \beta_{S} (\beta_{T1} \alpha_{V1}^2
-2 \alpha_{V2} \beta_{T2} \alpha_{V1}
-\alpha_{V2}^2 \beta_{T1})\bigg) \kappa_4
+3 x^2 \bigg((208 \beta_{S}^2+48 \beta_{T1}^2+48 \beta_{T2}^2+15(\beta_{V1}^2+\beta_{V2}^2)) \alpha_{S}^2+6 (48\alpha_{T1}\beta_{S} \beta_{T1}+25 \sqrt{3} \alpha_{V1} \beta_{V1} \beta_{T1}
+25 \sqrt{3} \alpha_{V2} \beta_{V2}\beta_{T1}
-99 \alpha_{V1} \beta_{S} \beta_{V1}
-25 \sqrt{3} \alpha_{V2} \beta_{T2} \beta_{V1}
+99 \alpha_{V2} \beta_{S}\beta_{V2}
+25 \sqrt{3} \alpha_{V1} \beta_{T2} \beta_{V2}
+6 \alpha_{T2} (\sqrt{3} \beta_{V1} \beta_{V2}
-8 \beta_{S}\beta_{T2})
+3 \sqrt{3} \alpha_{T1} (\beta_{V2}^2
-\beta_{V1}^2)) \alpha_{S}
+3 (16 \alpha_{T1}^2
+16 \alpha_{T2}^2
+5(\alpha_{V1}^2
+\alpha_{V2}^2)) \beta_{S}^2
+9 (4 (-4 \beta_{T1}^2
-4 \beta_{T2}^2
+\beta_{V1}^2
+\beta_{V2}^2) \alpha_{T1}^2
+24 \beta_{T1} (\alpha_{V2} \beta_{V2}
-\alpha_{V1} \beta_{V1}) \alpha_{T1}
+4 \alpha_{V1}^2\beta_{T1}^2
+4 \alpha_{V2}^2 \beta_{T1}^2
+4 \alpha_{V1}^2 \beta_{T2}^2
+4 \alpha_{V2}^2 \beta_{T2}^2
+19 \alpha_{V1}^2\beta_{V1}^2
+\alpha_{V2}^2 \beta_{V1}^2
+(\alpha_{V1}^2
+19 \alpha_{V2}^2) \beta_{V2}^2
-36 \alpha_{V1} \alpha_{V2} \beta_{V1}\beta_{V2}
+24 \alpha_{T2} \beta_{T2} (\alpha_{V1} \beta_{V1}
-\alpha_{V2} \beta_{V2})
+4 \alpha_{T2}^2 (
-4 \beta_{T1}^2
-4\beta_{T2}^2
+\beta_{V1}^2
+\beta_{V2}^2))
-6 \sqrt{3} \beta_{S} (3 \beta_{T1} \alpha_{V1}^2
+(-6 \alpha_{V2}\beta_{T2}
-25 \alpha_{T1} \beta_{V1}
+25 \alpha_{T2} \beta_{V2}) \alpha_{V1}
-\alpha_{V2} (3 \alpha_{V2} \beta_{T1}
+25 (\alpha_{T2}
   \beta_{V1}
+\alpha_{T1} \beta_{V2})))\bigg)\bigg)>0
\end{autobreak}
\end{align}

\begin{table}[H]\normalsize
    \centering
    \begin{tabular}{|c|c|c|c|c|}
    \hline
         $\alpha_{T1}$ $\beta_{T1}$&$\alpha_{T2}$ $\beta_{T2}$ & $\alpha_{V1}$ $\beta_{V1}$ & $\alpha_{V2}$ $\beta_{V2}$ & $\alpha_{S}$ $\beta_{S}$ \\
                  \hhline{|=|=|=|=|=|}
         $0,0$ &$0,0$& $0,0$ &$0,0$&$1,1$\\
         \hline
         $0,0$ &$0,0$& $1,1$ &$0,0$&$0,0$\\
         \hline
         $-0.100,-0.100$ &$0.750,-0.750$& $0,0$ &$0,0$&$0.654,0.654$\\
         \hline
         $-0.017,-0.017$ &$0.492,-0.492$& $-0.317,0.317$ &$0.306,0.306$&$0.751,0.751$\\
         \hline
         $-0.004,0.004$ &$-0.001,-0.001$& $-0.049,-0.049$ &$0.474,-0.474$&$0.879,-0.879$\\
         \hline
         $0,0$ &$0,0$& $0.939,-0.343$ &$0.343,-0.939$&$0,0$\\
         \hline
         $-0.508,0.508$ &$-0.196,-0.196$& $0.651,0.651$ &$0.121,-0.121$&$0.515,-0.515$\\
         \hline
        $-0.160,0.160$ &$-0.131,-0.131$& $-0.283,-0.283$ &$0.792,-0.792$&$0.499,-0.499$\\
         \hline
         $-0.426,0.426$ &$-0.046,-0.046$& $-0.044,-0.044$ &$0.819,-0.819$&$-0.379,0.379$\\
         \hline
         $0.042,0.042$ &$0.211,-0.211$& $-0.617,0.617$ &$-0.506,-0.506$&$-0.563,-0.563$\\
        \hline\hline
         $0.676,0.676$ &$-0.004,0.004$& $-0.002,0.002$ &$-0.572,-0.572$&$0.465,0.465$\\
         \hline
         $0.455,-0.455$ &$0.206,0.206$& $-0.171,-0.171$ &$0.791,-0.791$&$0.310,-0.310$\\
         \hline
         $0.795,0.795$ &$-0.046,0.046$& $-0.003,0.003$ &$-0.116,-0.116$&$0.593,0.593$\\
         \hline
         $-0.526,0.526$ &$0.521,0.521$& $-0.158,-0.158$ &$-0.383,0.383$&$-0.529,0.529$\\
         \hline
         $0.301,-0.301$ &$-0.532,-0.532$& $0.343,0.343$ &$0.589,-0.589$&$0.402,-0.402$\\
         \hline
         $0.362,-0.362$ &$0.145,0.145$& $0.715,0.715$ &$0.138,-0.138$&$-0.564,0.564$\\
         \hline

    \hline
    \end{tabular}
    \caption{ Special configurations of polarizations for $hh\rightarrow hh$ scatterings in SVT basis which give strong constraints on the couplings.
    These are determined by minimizing the $\alpha$'s and $\beta$'s for guessed Lagrangian parameters. This procedure is repeated until a compact parameter region is obtained. In the case of line theories, the last six configurations of polarizations are needed, in addition to the first ten, to give strong constraints on the couplings.}
    \label{tab:3}
\end{table}

\subsection{$hf\rightarrow hf$ Scattering Amplitude with Suppressed Cubic Interaction}
\begin{align}\label{indef bound 2}
f=&\frac{g_*^2}{\gamma \Lambda_2^4 x^2}\Bigg(\Bigg.\frac{ (4 \alpha_{S}^2+3 \alpha_{V1}^2+3 \alpha_{V2}^2) (4 \beta_{S}^2+3 \beta_{V1}^2+3 \beta_{V2}^2)}{12 }( \kappa^{(2)}_{3} c_{1}+ \kappa^{(1)}_{3} c_{2})\\
&+\frac{c_{2} (4 \beta_{S}^2+3 \beta_{V1}^2+3 \beta_{V2}^2) (2 \alpha_{S}^2+6 \alpha_{T1}^2+6 \alpha_{T2}^2+3 \alpha_{V1}^2+3 \alpha_{V2}^2)}{18 }\nn\\
&+\frac{c_{1} (4 \alpha_{S}^2+3 \alpha_{V1}^2+3 \alpha_{V2}^2) (2 \beta_{S}^2+6 \beta_{T1}^2+6 \beta_{T2}^2+3 \beta_{V1}^2+3 \beta_{V2}^2)}{18  }\nn\\
-&\Bigg(\Bigg.\alpha _{S}^2 (8 \beta _{S}^2+6 (\beta _{V1}^2+\beta _{V2}^2))-6 \alpha _{S}\bigg(4\beta _{S}( \alpha _{T1}\beta _{T1}- \alpha _{T2}\beta _{T2})+\sqrt{3} \alpha _{T1} (\beta _{V2}^2-\beta _{V1}^2)+2 \sqrt{3} \alpha _{T2} \beta _{V1} \beta _{V2}\bigg)\nn\\+&6 \beta _{S}^2 (\alpha _{V1}^2+\alpha _{V2}^2)+6 \sqrt{3} \beta _{S} ((\alpha _{V1}^2 -\alpha _{V2}^2) \beta _{T1}-2 \alpha _{V1} \alpha _{V2} \beta _{T2})+9 (\alpha _{V1} \beta _{V2}+\alpha _{V2} \beta _{V1})^2\Bigg. \Bigg)\frac{\lambda}{18 }\Bigg)>0.\nn
\end{align}

\bibliographystyle{JHEP}
\bibliography{references}

\end{document}